This manuscript is a preprint and has not yet been peer reviewed. Supporting information can be found after the references part

**Consistent transition model for $Bi_{0.5}Na_{0.5}TiO_3$ from temperature-dependent structural and electrical properties**

*Thomas Fourgassie [1,2], Omar Ibder [2], Cosme Milesi-Brault [2], Anna Katharina Ott [1], Eric Bourhis [3], Pascal Andreazza [3], Pierre-Eymeric Janolin [2], and Cécile Autret-Lambert * [1,2]*

Affiliations :

[1] Laboratoire GREMAN UMR7347, University of Tours, Tours, France

[2] Université Paris-Saclay, CentraleSupélec, CNRS, Laboratoire SPMS, Gif-sur-Yvette, France

[3] ICMN UMR7347, University of Orléans, Orléans, France

*Corresponding author e-mail: autret@univ-tours.fr*

**Abstract**

$Bi_{0.5}Na_{0.5}TiO_3$ (BNT)-based solid solutions are promising parent materials for lead-free dielectric capacitors, thanks to their high recoverable energy densities and breakdown strengths. However, the ambient-temperature symmetry and high-temperature phase evolution of BNT remain unclear. Crucially, structural transformations and electrical ordering are most often considered independently, hindering a coherent understanding of the BNT phase transition.

In this work, we combine X-ray diffraction (XRD), transmission electron microscopy (TEM), and Raman spectroscopy, impedance spectroscopy, and high-field polarization cycling to establish a unified picture of the structural and dielectric response of BNT. Based on these results, we propose a consistent transition model for BNT that reconciles previously conflicting interpretations. This integrated structure–property study provides a rationale for developing high-performance, lead-free energy-storage materials.

## 1. Introduction

$Bi_{0.5}Na_{0.5}TiO_3$ (BNT) is a lead-free perovskite material with ferroelectric properties that has sparked much interest due to its promising piezoelectric and dielectric properties for energy storage and energy conversion applications. [1–3] Despite extensive efforts, understanding its structure is still a challenge, even at room temperature, due to the coexistence of multiple phases with close symmetries and $Na^+/Bi^{3+}$ cation disorder on the A-site of the perovskite. [4,5]

Initially, the room-temperature structure of BNT was assumed to be rhombohedral with the *R*3*m* space group. [6] However, subsequent studies demonstrated that, while BNT was indeed rhombohedral, it belongs to the *R*3*c* space group [7] based on the oxygen octahedra tilts pattern observed by neutron diffraction ( $a^-a^-a^-$ in Glazer's notation [8]) that is only compatible with the *R*3*c* or $R\bar{3}c$ space groups. [9,10] Only the *R*3*c* space group is polar and therefore compatible with the ferroelectric response BNT exhibits [11]. Despite the wide consensus on the existence of the *R*3*c* phase at room temperature, a more detailed description has been advocated to account for the tetragonal platelets and local distortions. [12–16] Alternatively, the BNT low-temperature monoclinic *Cc* phase is also suggested for BNT at room temperature, [15] either as the single space group or coexisting with the *R*3*c* phase. This equilibrium between the two phases appears highly dependent on the material's history, such as the heat treatment or the applied electric field. [17–21] Although the room temperature structure of BNT remains controversial, its high-temperature phase evolution is even less understood. The most widely used transition model assumes that BNT has a rhombohedral *R*3*c* phase at room temperature [22–24] that is proposed to transit above the depolarization temperature ($T_d$, around 200 °C) in a tetragonal *P*4*bm* phase, creating an ergodic relaxor state due to the disordered coexistence of the *R*3*c* and *P*4*bm* phases. Then, above $T_m$ (around 320 °C), the sample enters a paraelectric state with a tetragonal (pseudo-cubic) pure *P*4*bm* structure, which becomes completely cubic around 520 °C. [25] While this transition model has been widely adopted, other models have been proposed. Notably, the Dorcet and Trolliard transition model. [12,13,26] This model puts forward a coexistence of the *R*3*c* and *P*4*bm* phases already at room temperature. At $T_d$, BNT enters a two-phase region comprising rhombohedral *R*3*c* and orthorhombic *Pnma* phases; the *R*3*c* phase disappears by 280 °C, leaving only the *Pnma* phase to persist over a narrow temperature interval between 280 °C and $T_m$ (at 320 °C). Above 320 °C, BNT transits to the pseudo-cubic paraelectric phase. The disadvantage of this model is that it was constructed solely from structural property measurements (mainly temperature-dependent transmission electron microscopy), resulting in a loss of information about the electrical order of the structure.

The structural complexity of the parent material BNT extends to BNT-based solid solutions *e.g.*, BNT-$BaTiO_3$ or BNT-$CaTiO_3$ [27–29], which are widely studied for their improved functional properties compared to pure BNT. Therefore, there is a need for a phase-transition model that can reconcile studies comparing BNT's structural and electrical properties. Indeed, most studies have focused primarily on structural analyses using X-ray diffraction (XRD), Raman spectroscopy, neutron or electron diffraction, without directly linking the structures to the observed dielectric properties (both at low and high fields). Our study bridges the gap between these two approaches and constructs a consistent phase-transition model for pure BNT, aiming to bring the community closer to consensus.

## 2. Results and discussions

### 2.1 Microstructure and chemical homogeneity

The chemical composition and microstructure of our BNT sample must be determined as a crucial prerequisite for characterizing its structural and electrical properties. **Figure 1** illustrates the homogeneous microstructure of the sample and its monodisperse grain-size distribution, as determined by scanning electron microscopy (SEM). The average grain size is 0.462 μm (with a standard deviation of 0.176 μm), indicating a sub-micrometric grain size. This average grain size is consistent with our previously reported [30] synthesis process, which uses both the gel-assisted citrate process and multi-step sintering.

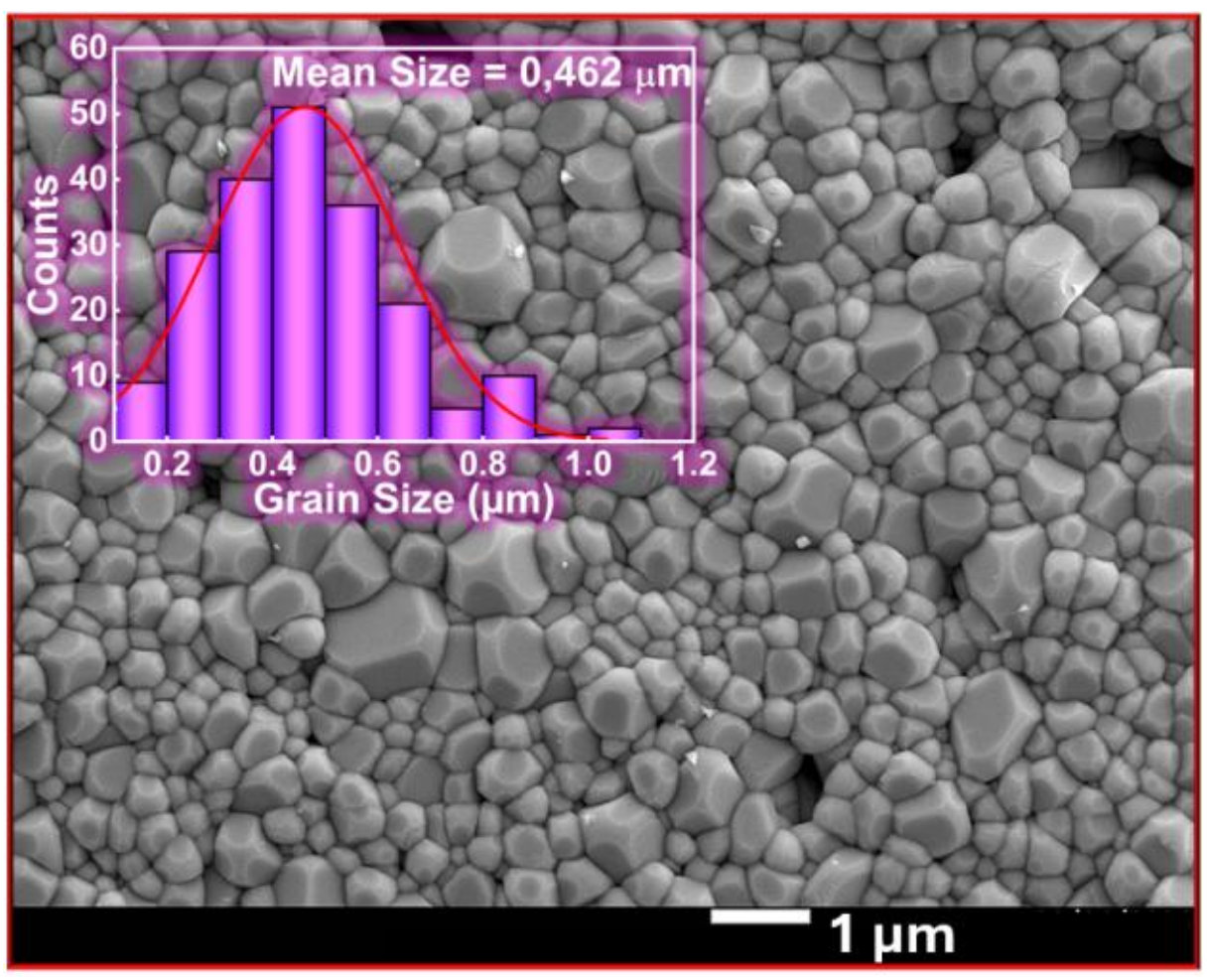

**Figure 1 –** Homogeneous microstructure of the BNT ceramic imaged by SEM.

Inset: grain-size distribution confirming a submicrometric average grain size.

In addition to microstructural homogeneity, the sample exhibits chemical homogeneity as observed in energy-dispersive X-ray (EDX) mapping (**Figure S1** in the supporting information**)**. Furthermore, the element quantification showed a chemical composition consistent with the target composition of $Bi_{0.5}Na_{0.5}TiO_3$, as shown in **Figure S2**, taking into account the instrument error for cation quantifications (~0.02) and the larger error for oxygen, which is always overestimated by the EDX detector.

To further ascertain the composition at room temperature, X-ray photoelectron spectroscopy (XPS) was also performed. If the composition differed from the target composition, we would expect to observe mixed oxidation states in the XPS spectrum, particularly for titanium ions, which tend to transition from $Ti^{4+}$ to $Ti^{3+}$. **Figure 2** shows a zoomed region of the resulting XPS spectrum around the three characteristic peaks of the three cations (Na, Bi, and Ti). For bismuth, the 4*f* peaks (**Figure 2.a**) could be fitted perfectly using only one doublet corresponding to the $4f_{7/2}$ peak at a binding energy of 158.25 eV and the $4f_{5/2}$ peak at 163.55 eV. [31] For titanium, the 2*p* peak (**Figure 2.b**) was also successfully fitted using only one doublet, corresponding to the $2p_{3/2}$ peak at 457.45 eV and the $2p_{1/2}$ at 463.15 eV that overlaps with one of the 4*d* peaks of bismuth. Apart from this overlap, no shoulder is observed on the titanium 2*p* doublet, indicating that the titanium has a unique oxidation state. Although the binding energy of the doublet differs slightly from the one expected for $Ti^{4+}$, the 5.7 eV gap between the peaks of the doublet can only correspond to a 4+ oxidation state and not to a 3+ state. [32,33] In the case of the sodium cation, the Na 1*s* peak (**Figure 2.c**) was perfectly fitted with one singlet at 1070.95 eV, as expected for $Na^+$ ions and confirming the single oxidation state. The two other visible peaks are attributed to LMM Auger peaks due to the titanium.

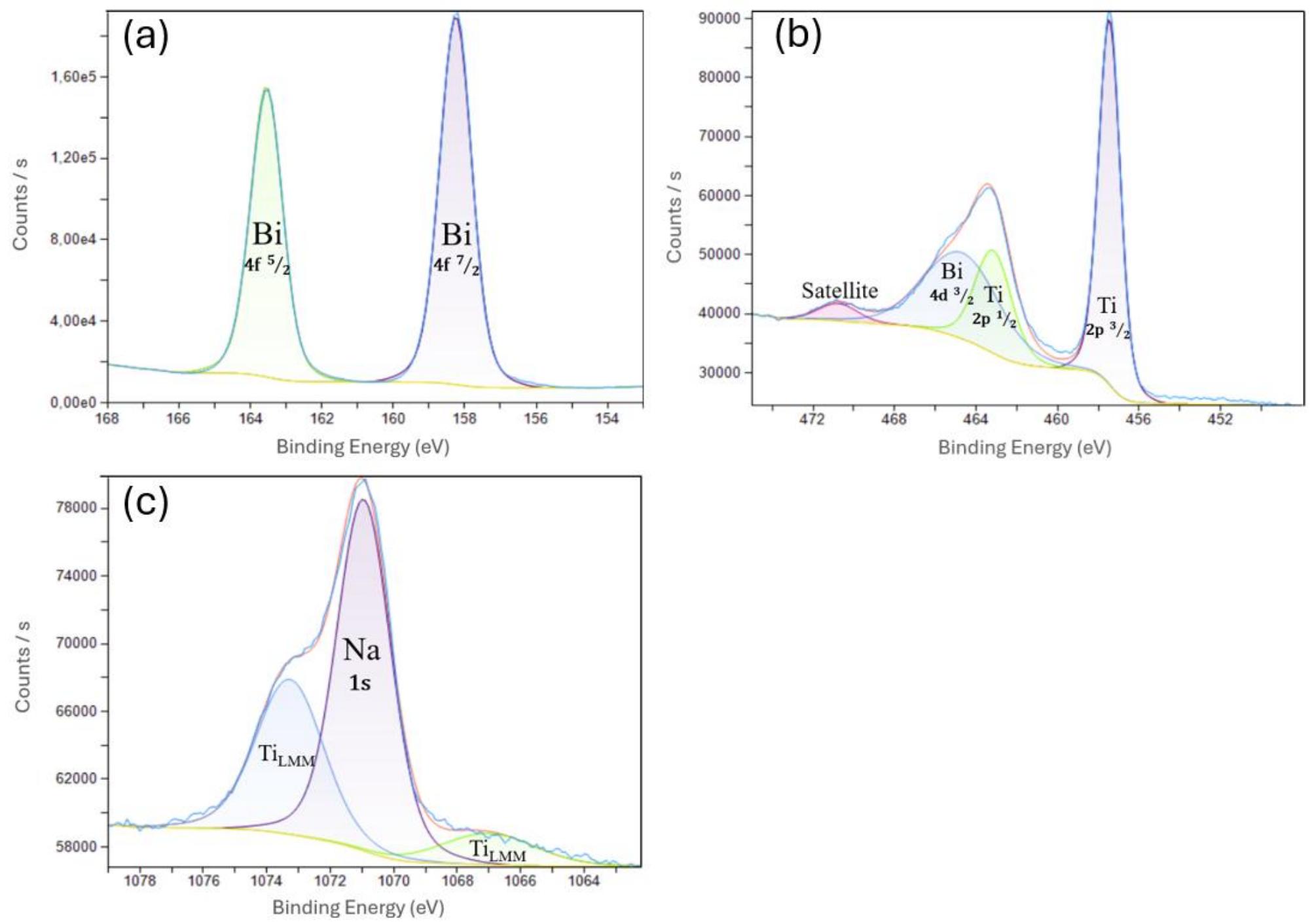

**Figure 2** – Confirmation of the expected chemical composition of the BNT sample from XPS peak fitting for each cation, showing a single oxidation state for each element.

SEM/EDX and XPS analyses confirmed that the BNT sample is homogeneous, both in microstructure and in chemical composition. Therefore, the sample possesses the expected chemical composition.

**2.2 Structure at room temperature**

The X-ray diagram was initially refined using the $R3c$ space group, yield cell parameter values a = 5.488885(6) Å and c ≈ 13.4685(2) Å (V ≈ 351.41(8) $Å^3$) in good agreement with those reported previously. [23] Peak profiles were fitted using a pseudo-Voigt function. Although all reflections were indexed, the initial Rietveld refinement showed a poor agreement between the observed and calculated patterns (**Figure S3.a**). The reliability factors for this refinement were $R_p$ = 7.99 %, $R_{wp}$ = 10.3 %, $R_{Bragg}$ = 4.88 % and $\chi^2$ = 4.44. The isotropic atomic displacement parameters ($B_{iso}$) were very high (>4) for the A-site cations, while the occupation factors remain consistent with the nominal composition, indicating disorder between Bi and Na cations. This disorder was therefore modeled using anisotropic displacement parameters.

A detailed inspection of the diffraction pattern reveals additional discrepancies (**Figure S3**). The intensities of the $(0\ k\ 2k)_H$ reflections, where H stands for hexagonal indexation, are significantly reduced (**Figure S3.d**). Some peaks are broadened, and a shoulder is observed on the high-angle side of the $(018/214/300)_H$ peaks (marked by a star in **Figure S3.e**). This shoulder can be attributed to the *P*4*bm* tetragonal phase (a ≈ 5.50 Å, c ≈ 3.91 Å, and V ≈ 118 $Å^3$) [12,14] from the Dorcet and Trolliard model. An alternative explanation for the broadened peak and decreased intensity is the presence of microstrains induced by lattice defects, compositional inhomogeneities, or internal stresses, which produce non-uniform crystal distortions. [34] While broadening occurs for several reflections, the $(0\ k\ 2k)_H$ interplanar spacing $d_{hkl}$ is altered, inducing angular shifts of the corresponding Bragg reflections ($\Delta 2\theta = -2\varepsilon\tan\theta$, where $\varepsilon = \Delta d/d$ is the relative variation of interplanar spacing). These effects may originate from the tetragonal phase and/or A-site disorder.

Considering these effects, the final refinement includes a tetragonal phase, relative variations in the $d_{0k2k}$ interplanar spacing, and microstrains. It provides a good fit to the experimental data (**Figure S4**), with $R_p$ = 6.42 %, $R_{wp}$ = 8.42 %, $R_{Bragg}$ = 3.38 % for the main phase (*R3c*) and 8.42 % for the second phase (*P*4*bm*), and $\chi^2 \approx 1.31$. The lattice parameters are a = 5.48997(4) Å, c = 13.4674(1) Å, and V = 351.484(4) $Å^3$ for the rhombohedral phase and a = 5.5024(2) Å, c = 3.9072(2) Å, and V = 118.30(1) $Å^3$ for the tetragonal phase.

Thus, the room-temperature XRD pattern of BNT can be refined using the model from Dorcet and Trolliard [12], a two-phase model comprising both rhombohedral *R3c* and tetragonal *P4bm* phases, with a microstrain model to account for defects (**Figure 3**). The tetragonal phase accounts for approximately 6.5% of the total structure. This low volumetric fraction explains why many satisfactory refinements using the *R3c* phase alone have been reported. [35–37]

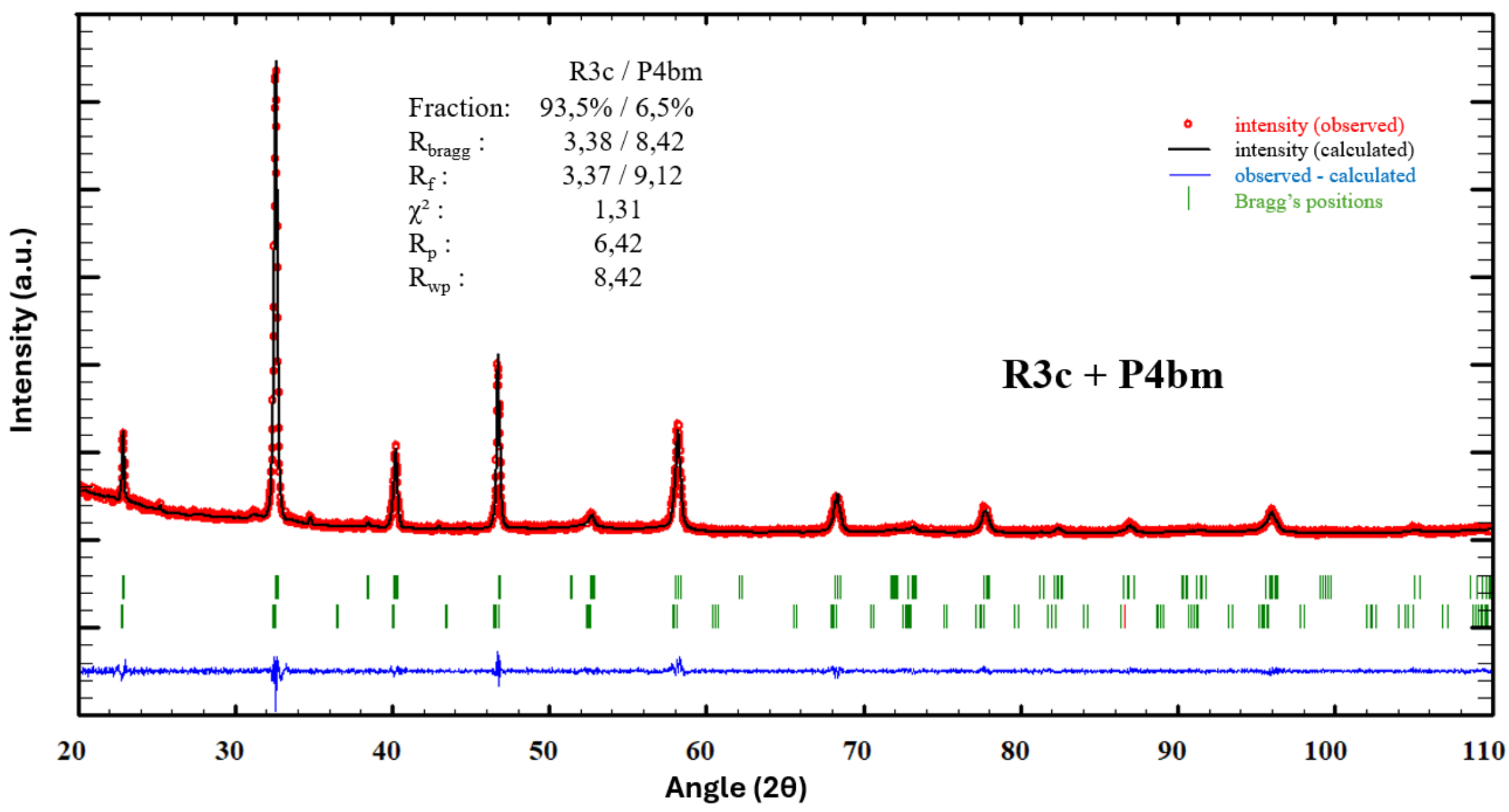

**Figure 3** – Rietveld refinement of the room temperature XRD pattern of BNT considering the coexistence of *R*3*c* and *P*4*bm* phases.

Thus, XRD analyses suggest the presence of both *R*3*c* and *P*4*bm* symmetries. To confirm the presence of the tetragonal phase without ambiguity, electron diffraction (ED) experiments were performed. This technique is particularly well-suited to determine the rotation modes of the oxygen octahedra from additional (superstructure) spots in the diffraction pattern, compared to the cubic perovskite structure [38]. Indeed, the oxygen octahedra tilt systems of the *R*3*c* and *P*4*bm* phases ($a^-a^-a^-$ for the former and $a^0a^0c^+$ for the latter, in Glazer notation [8]) generate different superstructure spots.

Two specific zone-axes ($[011]_{PC}$ and $[001]_{PC}$ in pseudo-cubic indexation) are particularly suited to reveal the characteristic features of each phase. Observation along the $[011]_{PC}$ axis reveals additional spots due to rhombohedral distortions and associated octahedral tilting modes. In contrast, the $[001]_{PC}$ pattern is useful for demonstrating tetragonal symmetry. Therefore, combined analyses of these two orientations enable clear and unambiguous differentiation between the two structures. Consequently, numerous crystallites have been characterized by ED, EDS, and High-Resolution Electron Microscopy (HREM). **Figure 4** exhibits ED patterns of BNT along the $[011]_{PC}$ and $[001]_{PC}$ zone axes. On the $[011]_{PC}$ pattern, in addition to the intense reflections consistent with the pseudo-cubic structure, weaker reflections, indicated by the red circles, are observed. These reflections correspond to the ½(ooo) superstructure reflections (where "o" stands for odd Miller indices), which arise from $a^-a^-a^-$ antiphase octahedral tilting around the $[111]_{PC}$ direction. The presence of these superstructure reflections provides clear evidence for the *R3c* phase, as they are forbidden in the *P4bm* phase. On the other hand, on the $[001]_{PC}$ pattern, the ½ (ooe) (where "e" stands for even Miller indices) superstructure reflections can be observed, noted by a green circle, which are characteristic of the *P4bm* phase and forbidden for the *R3c* phase. [39,40]

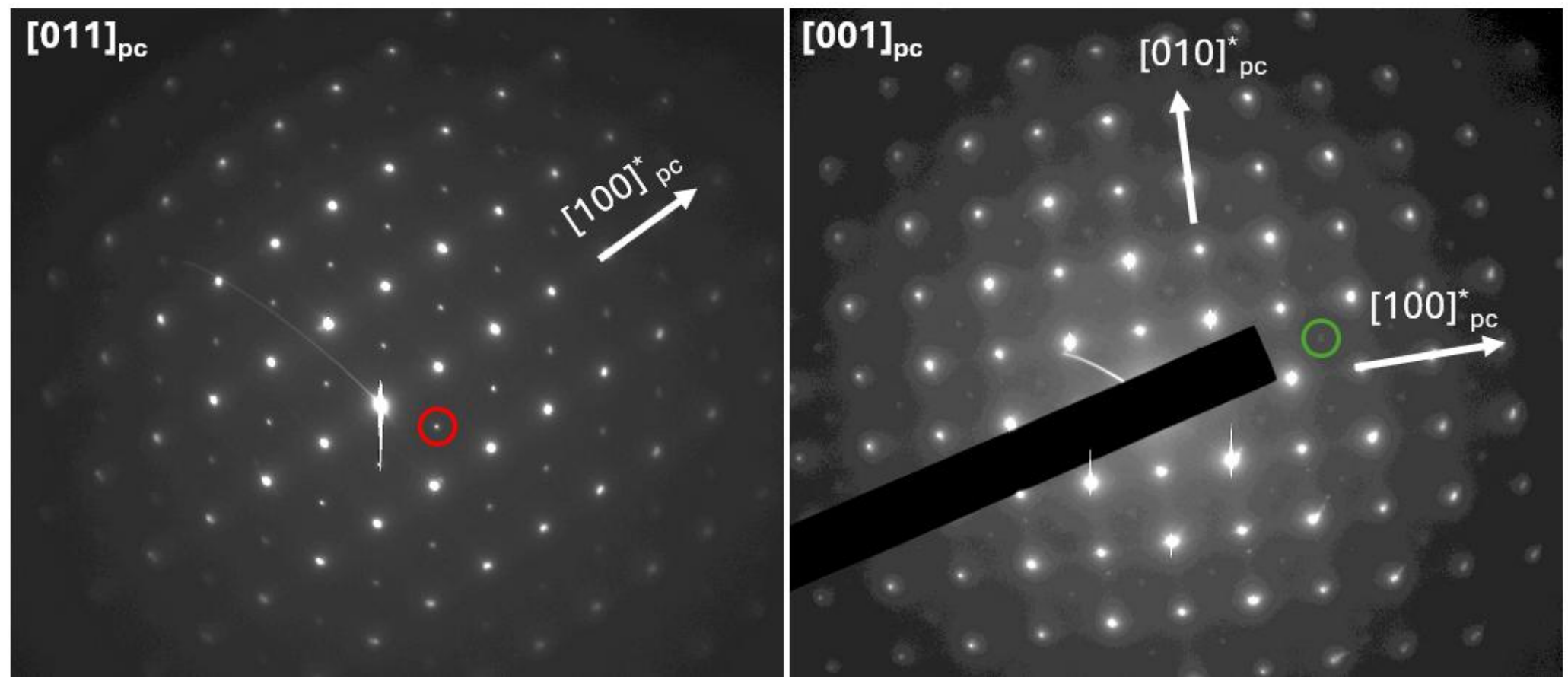

**Figure 4 –** Coexistence of the *R3c* and *Pb4m* phases evidenced from electron diffraction patterns of BNT along the $[011]_{PC}$ and $[001]_{PC}$ zone axes. The $[011]_{PC}$ zone axis shows ½ (ooo) superstructure reflections (red circle) characteristic of the *R3c* phase, while the $[001]_{PC}$ pattern shows ½ (ooe) reflections allowed by the *P4bm* phase.

Thus, the ED study shows that the *R*3c and *Pb*4*m* phases coexist in the sample at room temperature. The very weak intensity of the ½ (ooe) reflections confirms that the tetragonal *Pb*4*m* phase is a minority phase, consistent with the XRD results.

Furthermore, diffuse streaks appear parallel to a* (see **Figure S5**) along the $[0\ k\ 2k]_H$ direction, equivalent to the $[100]_{PC}$ direction. These streaks can be interpreted as a signature of local structural defects, such as a disorder on the cationic site leading to microstrains in the lattice. This interpretation is consistent with the XRD results, which showed a slight shift of the $(0\ k\ 2k)_H$ reflections, indicative of lattice distortions along this direction, and evidenced by the disorder of the Bi and Na cations on the A-site. TEM imaging reveals the presence of stacking faults and "nano-twinning", providing direct microstructural evidence of these local defects.

To further investigate the local order and the phase distribution in the compound, a HRTEM study was performed (see **Figure 5**). Local Fourier transform (**Figures 5.b-d**) was performed on the areas outlined in **Figure 5.a,** revealing weak additional reflections, characteristic of the tetragonal phase and oriented along one of the three main axes of the pseudo-cubic cell: $[100]_{PC}$, $[010]_{PC}$, or $[001]_{PC}$, in agreement with previous reports [12]. The presence of these nanometric domains is further confirmed by the global Fourier transform (**Figure 5.e**), where the superposition of superstructure reflections along all three directions is clearly visible. The identification of these nanostructured *P*4*bm* variants provides direct structural evidence for the presence of tetragonal polar nanoregions (PNRs).

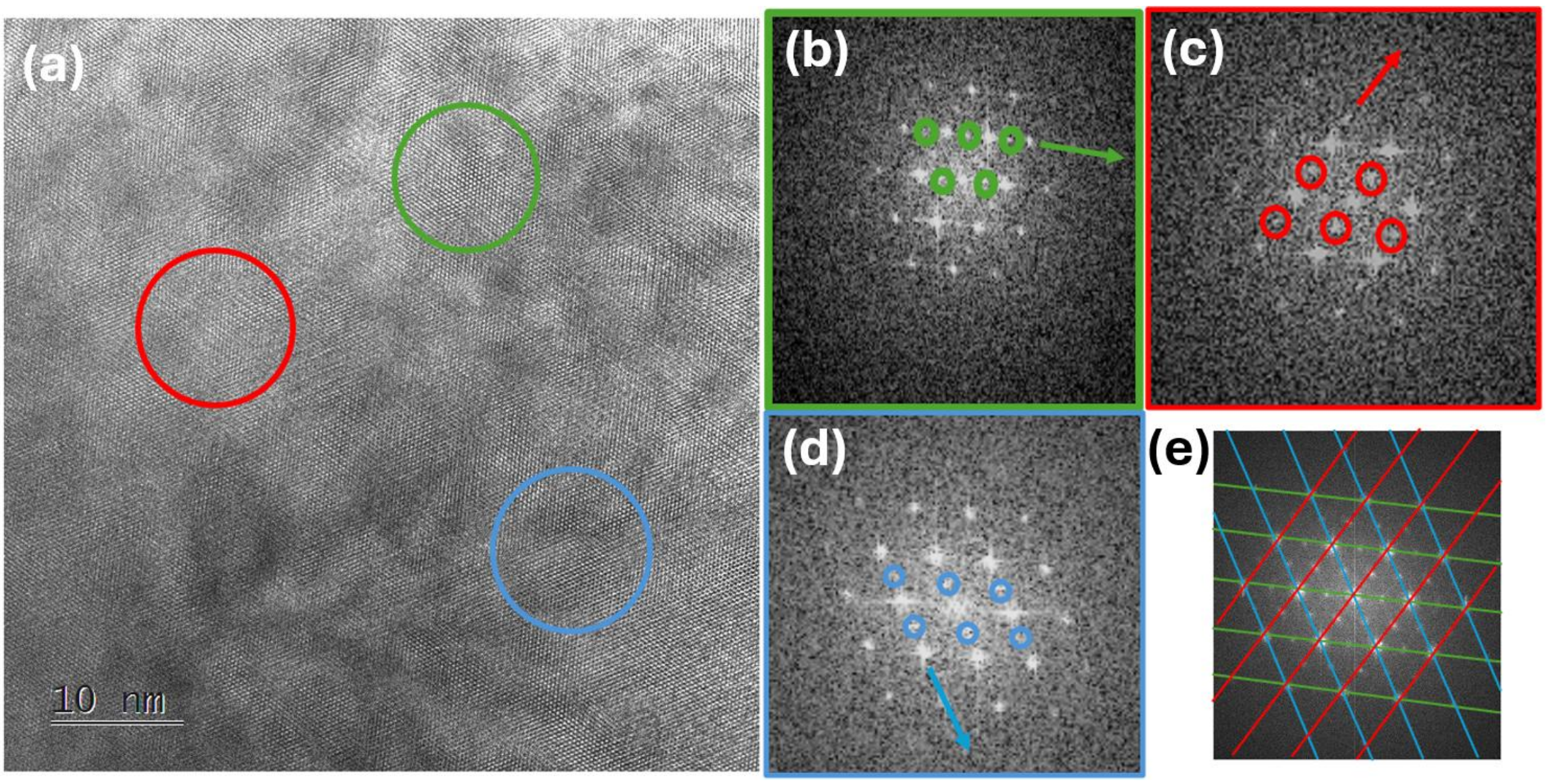

**Figure 5 –** (a) HRTEM image of BNT and corresponding Fourier transforms of selected regions (b–d), showing *P*4*bm* superstructure reflections along the <100>$_{PC}$ directions. (e) Fourier transform of the whole image showing the superposition of all orientations.

Under continuous exposure to the electron beam (200 kV accelerating voltage), the diffraction contrast from the nanodomains is enhanced, and their boundaries are more clearly defined (**Figure S6**). This phenomenon should not be interpreted as a degradation of the specimen during irradiation, but rather as an intrinsic response of the tetragonal nanodomains [41]. The energy delivered by the incident electron beam facilitates strain relaxation within the domains.

### 2.3 Temperature dependent transitions

#### 2.3.1 The depolarization temperature

The first transition in pure BNT generally occurs at the so-called depolarization temperature ($T_d$), typically reported between 160 °C and 200 °C. [30,42,43] Since the depolarization temperature is mainly a transition in polarization order, a significant change is expected when measuring polarization at high electric fields and crossing the $T_d$. Large changes in the temperature evolution of polarization are presented at selected temperatures in **Figure 6**. While only one current peak (for each field polarity) should be visible for a pure ferroelectric material, we observe a splitting of the current peak at 100 °C (**Figure 6.b**). Then, one of these two peaks crosses the zero-field line at 150 °C (**Figure 6.c**), accompanied by a significant reduction of the remanent polarization (from around 27 μC/cm² at 100 °C to around 12 μC/cm² at 150 °C), thereby defining the depolarization temperature. At 200 °C, a pinched hysteresis (antiferroelectric-like) cycle is observed (**Figure 6.d**).

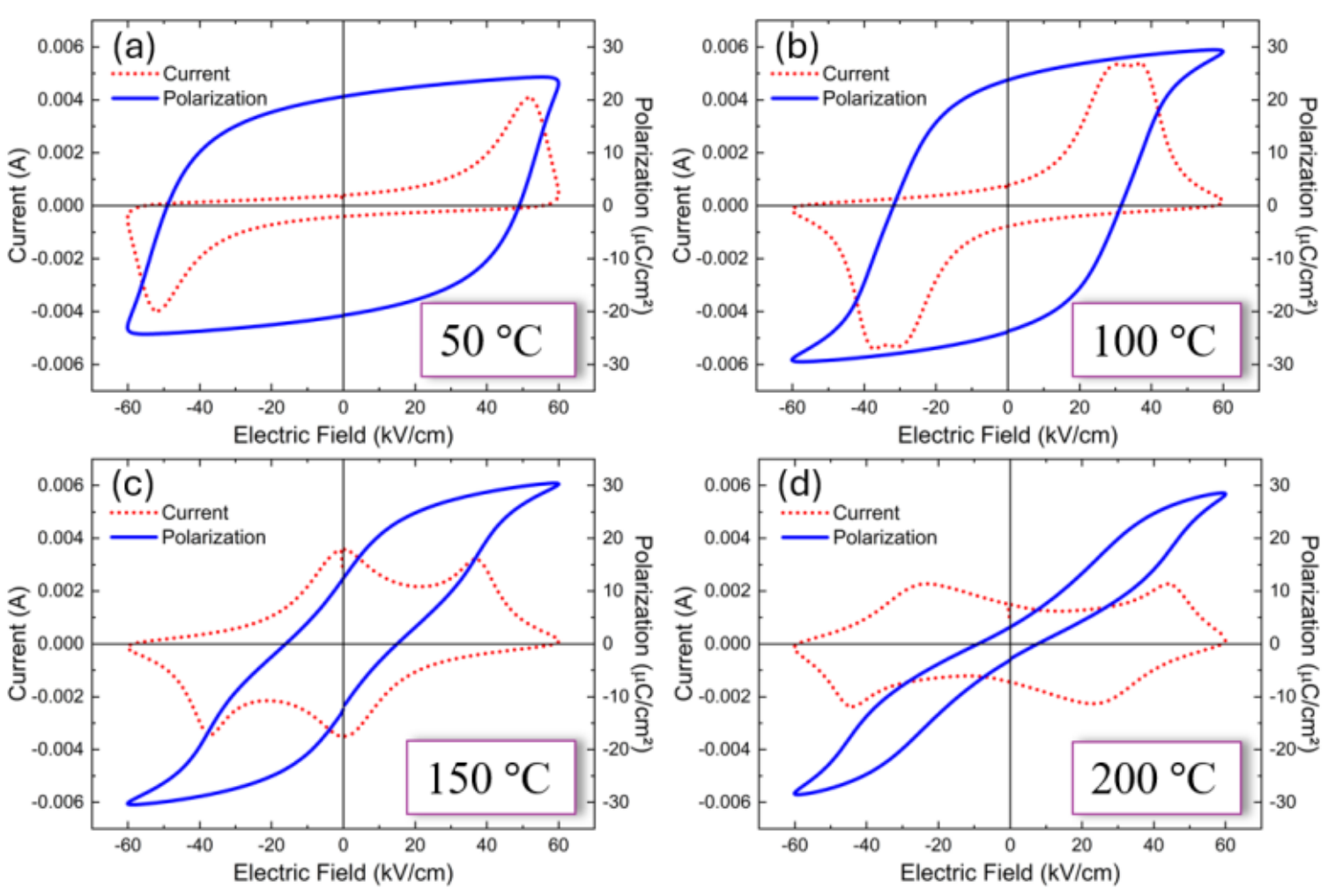

**Figure 6 –** Polarization cycles and current loops of BNT at (a) 50 °C, (b) 100 °C, (c) 150 °C and (d) 200 °C, showing current peak splitting and the emergence of an antiferroelectric-like response.

To confirm the depolarization temperature of 150 °C and clarify the origin of the transition, current loops were measured every 10 °C (**Figure 7.a**). Three different behaviors were identified: first, between 30 °C and 80 °C (**Figure 7.b**), there is one current peak for each field polarity, consistent with a pure ferroelectric response. The peaks move towards lower (absolute) electric field values, and their intensity progressively increases with temperature. This evolution is explained by the fact that, at room temperature, an electric field of 60 kV/cm is insufficient to fully saturate the sample's polarization. As the temperature increases, the coercive field decreases due to enhanced domain-wall mobility, thereby increasing the domain population responding to the field. [44,45] Second (**Figure 7.c**), between 80 °C and 150 °C, the intensity of the current peaks continuously decreases, with the splitting of each peak into two components. One of these components exhibits a coercive field that remains essentially independent of temperature, while the coercive field of the second component decreases progressively and reaches 0 kV/cm at 150 °C. The existence of this second component can be linked to the coexistence of the *P*4*bm* and *R*3*c* phases, the deconvolution of the seemingly single peak becoming possible only at higher temperatures because the coercive fields of the *R*3*c* and *P*4*bm* phases evolve at different rates. Above 150 °C (**Figure 7d**), the second component crosses over to the opposite polarity of the electric field axes, indicating the appearance of an antiferroelectric-like state. This confirms that 150°C corresponds to the depolarization temperature $T_d$. However, a broad distribution of coercive fields is observed, as expected for a complex material like BNT with coexisting structural phases. This distribution refutes the possibility of a pure antiferroelectric state. The non-zero remanent polarization (observed even at 200 °C, see **Figure 7.d**) is thus due to a persistent ferroelectric state.

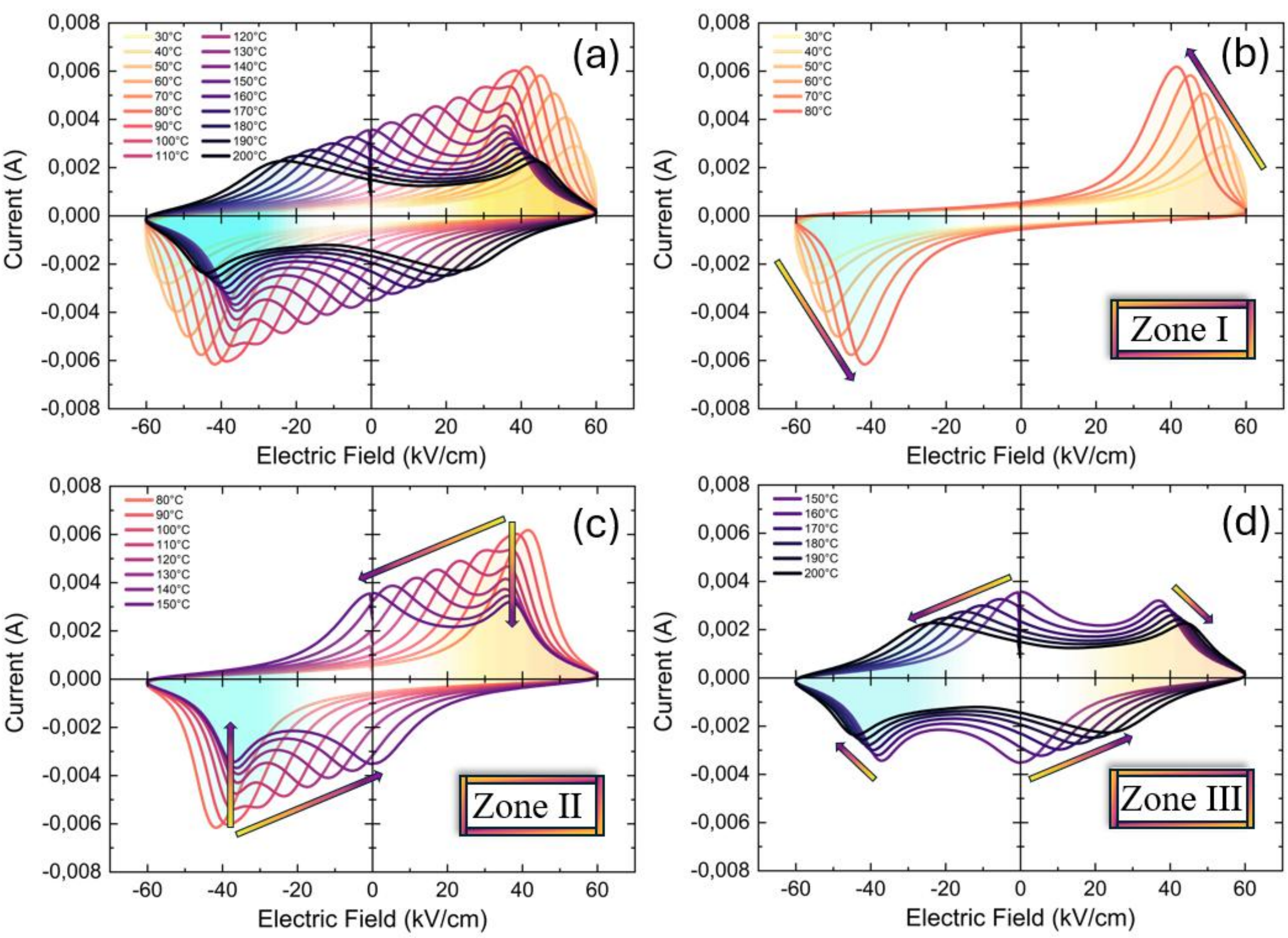

**Figure 7 –** Temperature evolution of the electric current : (a) global view from 30 °C and 200 °C, (b) single peak displacement between 30 °C and 80 °C, (c) peak splitting and shift toward lower fields of one component up to 150 °C, (d) between 150 °C and 200 °C development of an antiferroelectric-like state. Arrows indicate peak displacements with increasing temperature.

The main problem arising from such broad current peaks and the mixed electrical order is the complexity of separating the contributions of the different mechanisms. Therefore, the current peaks were fitted to quantify the evolution more precisely (**Figure S7**). Due to the highly symmetric nature of the current loops, only positive current values were fitted, using a current model in which the coercive field distribution is described by a pure Lorentzian function (see **Figure S7**). The peaks between 30°C and 70°C are well described by a single Lorentzian, but the sample is not yet fully saturated. At 80 °C, the fits reveal a clear splitting of the current peak into three distinct components (I1, I2, and I3). This splitting arises from the coexistence of the two different crystallographic structures in the sample, the rhombohedral *R*3*c* phase represented by the I1 current peak and the tetragonal *P*4*bm* phase, represented by two current peaks: I2 and I3. The fact that those two peaks belong to the same structure is supported by their temperature evolution, which shows they follow a similar linear decrease in the coercive field with increasing temperature (**Figure S7.d**). It can be inferred that the bimodal distribution of coercive field observed for the *P*4*bm* phase comes from the disorder caused by the orientation of the tetragonal platelets (as already observed with TEM), where a preferred orientation of the tetragonal platelets may possess a different coercive field distribution than the other orientations. The appearance of a further component (labeled I4) at 150 °C (resulting from the splitting of the I1 component, see **Figure S7.b**, appears at the same temperature where I2 crosses to opposite electric-field polarity, strongly suggesting the onset of an antiferroelectric response, whereas I1 is still the response of a ferroelectric contribution. Consequently, in BNT, the depolarization temperature $T_d$ would be assigned to the formation of this antiferroelectric state (I2/I4) coexisting with the *R3c* ferroelectric phase (I1). After 170 °C, the second current peak related to the *Pb*4*m* phase (I3) crosses to opposite electric-field polarity, generating a fifth current peak (I5). This corresponds to the other orientations of the *P*4*bm* phase starting to turn into an antiferroelectric phase. **Figure 8** shows the polarization loops obtained with the fitting of the current peaks at the important temperatures (80 °C, 150 °C, and 170 °C), showing both individual contributions of the different phases as well as the global fitted polarization. The comparison between the fitted loops and the measured ones shows a good fit between our model of current peak distribution and the experimental values.

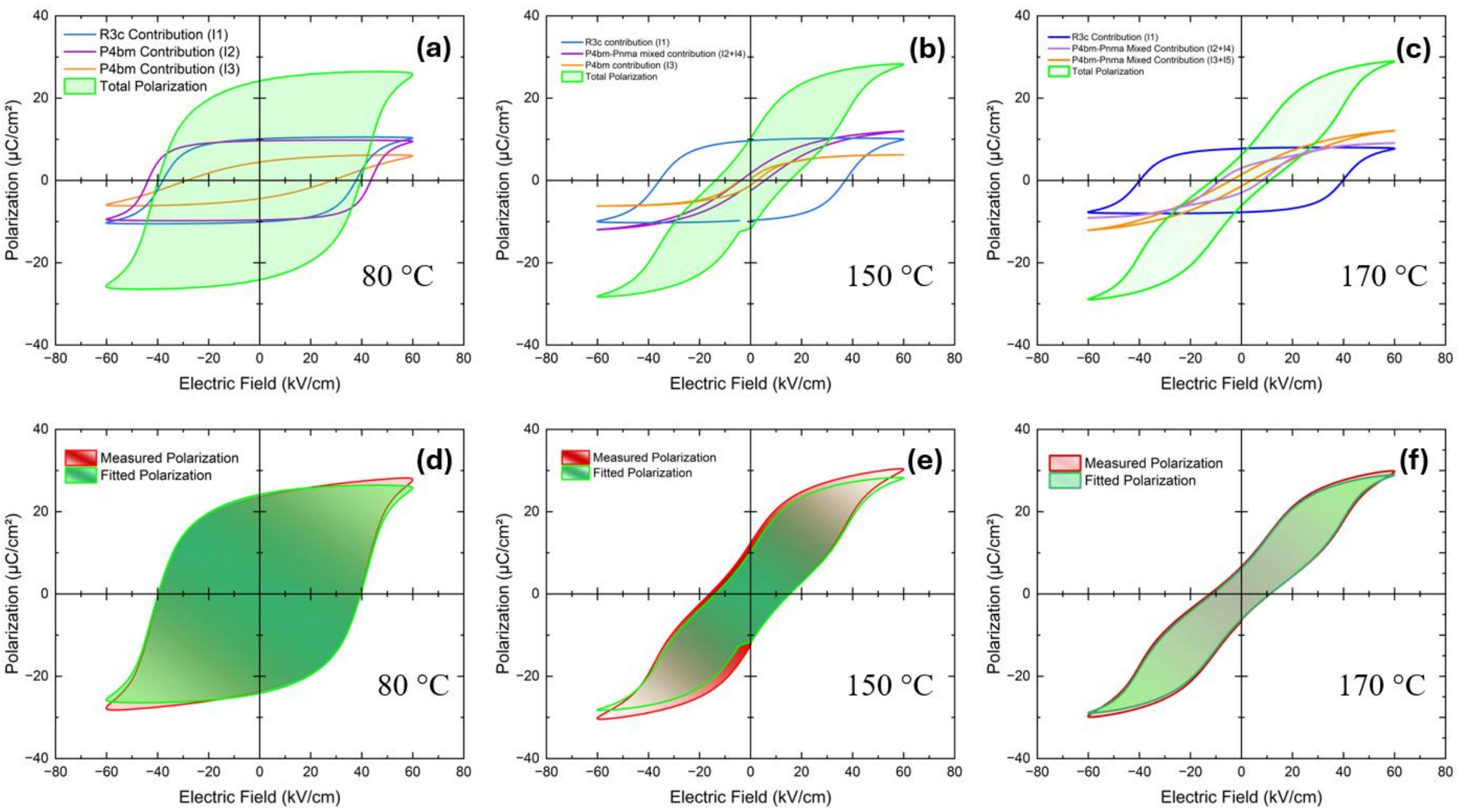

**Figure 8 –** Reconstruction of the polarization loops using the fitted curves of **Figure S7** showing the different contribution and the sum of those contributions at (a) 80 °C, (b) 150 °C and (c) 170 °C. Comparison of the reconstructed cycle with the measured polarization for (d) 80 °C, (e) 150 °C and (f) 170 °C.

This result is at odds with the commonly accepted model in which the *R*3*c* phase transitions to a pure *P*4*bm* phase at $T_d$. On the other hand, it is compatible with the scenario proposed by V. Dorcet and G. Trolliard, [13] in which a modulated phase forms: orthorhombic *Pnma* sheets appear between the *R*3*c* blocks, progressively creating an antiferroelectric-like order. To test this hypothesis, we compared the Rietveld refinement of the room-temperature structure (*R*3*c* + *P*4*bm*) with the model in which *P*4*bm* is replaced by *Pnma* phase while the *R*3*c* phase remains (*R*3*c* + *Pnma*) at 175 °C, just above $T_d$ (**Figure S8**). Both models give acceptable reliability factors, but the *R*3*c* + *Pnma* model provides slightly better agreement with the experimental data.

Combining the XRD results with analyses of the current loops, the most consistent hypothesis is that the depolarization temperature $T_d$ in BNT corresponds to a transition from the *P*4*bm* phase to the orthorhombic *Pnma* phase, which exhibits an antiferroelectric-like current response consistent with its symmetry.

### 2.3.2 The rhombohedral-orthorhombic transition and $T_m$

Such a transition (from *R*3*c*+*P*4*bm* to *R*3*c*+*Pnma*) is also reflected in the dielectric response, where the first anomaly associated with $T_d$ appears, although it remains broad and difficult to see accurately in an unpoled sample (**Figure S9**). The dielectric data also reveal $T_m$, the temperature of maximum permittivity, which is around 370 °C and corresponds to the transition to a paraelectric pseudocubic phase. However, according to the Dorcet and Trolliard model a third transition temperature should be seen between $T_d$ and $T_m$, this transition should be caused by the disappearance of the *R*3*c* phase leaving a pure *Pnma* phase (this transition from rhombohedral + orthorhombic to pure orthorhombic will be noted $T_{R\text{-}O}$).To support this hypothesis, XRD was performed between $T_d$ and $T_m$, revealing the disappearance of the $(113)_H$ peak above 250 °C (**Figure 9**) that is characteristic of the rhombohedral phase *R*3*c*. The disappearance of the $(113)_H$ peak at this temperature would imply that the *R*3*c* and *Pnma* coexistence is replaced by the single *Pnma* phase between 250 °C and 300 °C. The difficulty is that the complete disappearance of the *R*3*c* phase should produce a clear dielectric signature, as the *R3c* phase is the only ferroelectric phase above 250 °C. Instead, the dielectric response between 250 °C and 300 °C only shows a weak anomaly, preventing us from readily confirming the presence of such a transition (**Figure S9**).

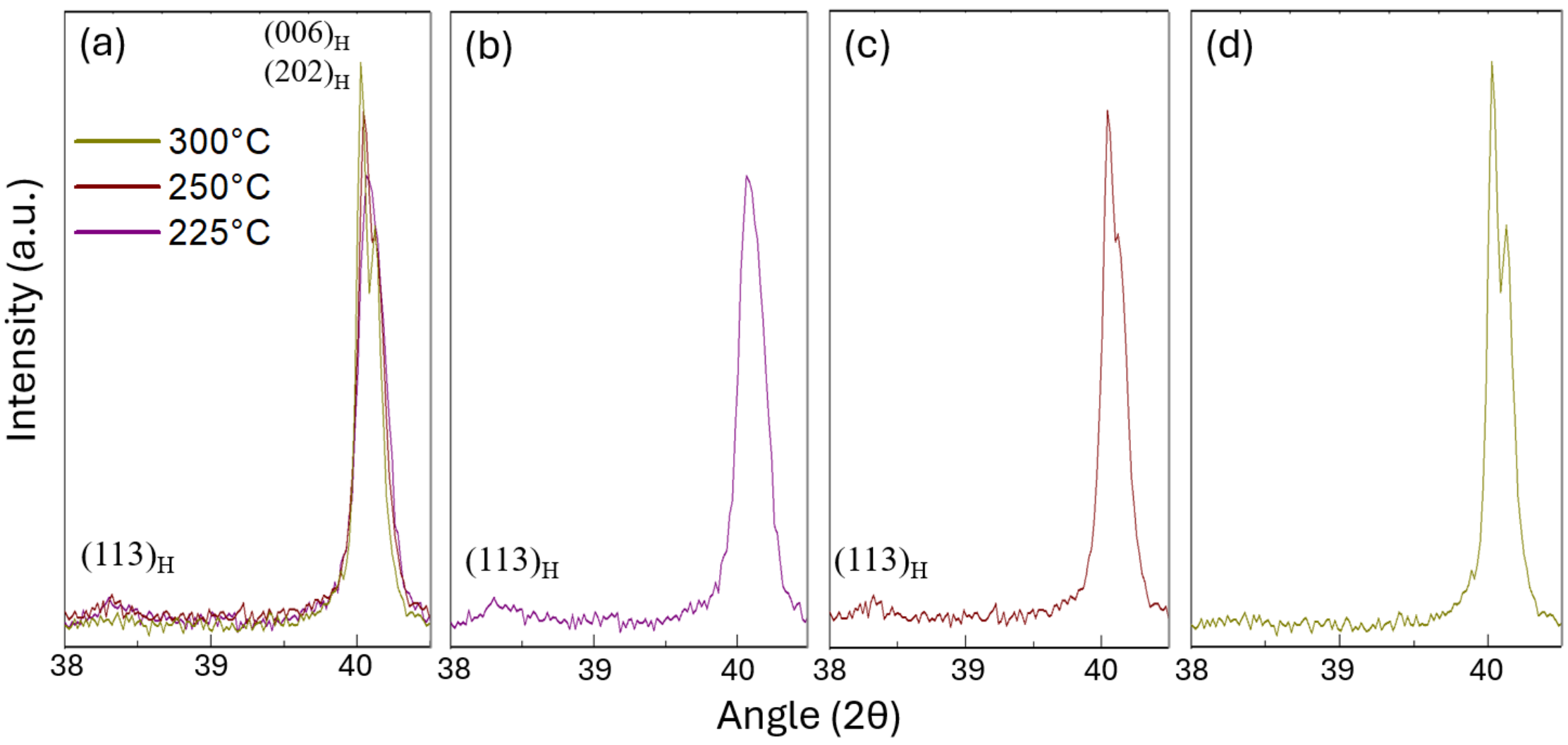

**Figure 9 –** (a) XRD patterns of BNT at elevated temperatures: (a) superimposed scans at 225, 250, and 300 °C; individual scans at (b) 225 °C, (c) 250 °C, and (d) 300 °C. The $(113)_H$ superstructure peak of the *R*3*c* phase disappears above 250°C.

The permittivity curves shown in **Figure S9** are directly provided by the impedance analyzer, which converts the raw impedance data using a default equivalent-circuit model, the "C" (for capacitor) model. However, this conversion inevitably results in a loss of information about the actual physical mechanisms at play. To address this limitation, the raw impedance data were re-examined using five different equivalent-circuit models (**Figure S10**). In addition to the C model, we considered the RC (resistance and capacitance) and RCPE (resistance and constant phase element) models in parallel configuration, followed by their respective series extensions, namely the 2RC and 2RCPE models. These latter two models allow the characterization of the grain and grain-boundary contributions.

The impedances of the C, RC, and 2RC models are given by **equations 1-3:**

$$Z_C = \frac{1}{j\omega C^*} \quad (1)$$

$$Z_{RC} = \frac{1}{\frac{1}{R} + j\omega C} \quad (2)$$

$$Z_{2RC} = \frac{1}{\frac{1}{R_1} + j\omega C_1} + \frac{1}{\frac{1}{R_2} + j\omega C_2} \quad (3)$$

Where $C^*$ is the complex capacitance, $\omega$ is the angular frequency (rad.s$^{-1}$), $C$ is a (real valued) capacitance (F), and $R$ is the resistance (Ω). For the RCPE and 2RCPE models, the impedance of an RCPE element is usually described by:

$$Z_{RCPE} = \frac{R}{1+Q_0(j\omega)^n} \quad (4)$$

With $Q_0$ the characteristic CPE parameter and $n$ ($0 \leq n \leq 1$) a parameter indicating the deviation from an ideal capacitor. However, **equation 4** does not provide a direct relation to an effective capacitance. Using the approach proposed by Byoung-Yong Chang, [46], a modified form can be used (**equation 5**) to fit the data for the RCPE model. The extension to the 2RCPE model results in **equation 6**.

$$Z_{RCPE} = \frac{R_{eff}\sin\left(n\frac{\pi}{2}\right)}{(1+j\omega R_{eff}C_{eff})^n} \quad (5)$$

$$Z_{2RCPE} = \frac{R_{eff1}\sin\left(n_1\frac{\pi}{2}\right)}{(1+j\omega R_{eff1}C_{eff1})^{n_1}} + \frac{R_{eff2}\sin\left(n_2\frac{\pi}{2}\right)}{(1+j\omega R_{eff2}C_{eff2})^{n_2}} \quad (6)$$

Equations 5 and 6 allow the impedance to be expressed in terms of effective parameters ($R_{eff}$, $C_{eff}$) that are more suitable for physical interpretation.

The Nyquist plots fitted at room temperature with the different models are presented in **Figure S11**. The RC and 2RC circuit models do not provide an adequate fit to the data, consistent with the fact that the dielectric losses of a highly disordered material such as BNT cannot be modeled as an ideal capacitor in series with a resistance.

The C model can be discarded based on its residuals (**Figure S12**), leaving the RCPE and 2RCPE as potentially appropriate models. The final selection of the most suitable model is based on information criteria (Akaike[47] and Bayesian[48,49]) that balance the model complexity with the agreement with the data (see Suppl. Info.). The statistical analysis of the models results in selecting the RCPE model.

The temperature evolution of the capacitance obtained with this model is shown in **Figure 10**. A hump is observed between 250 °C and 300 °C (circled in green), corresponding to the temperature range where the *R*3*c* phase disappears in favor of the *Pnma* phase in the X-ray diagrams (**Figure 9**). The use of an appropriate dielectric model enables evidence of the phase transition associated with the disappearance of the *R*3*c* phase, although no frequency dispersion is observed (see also **Figure S9).**

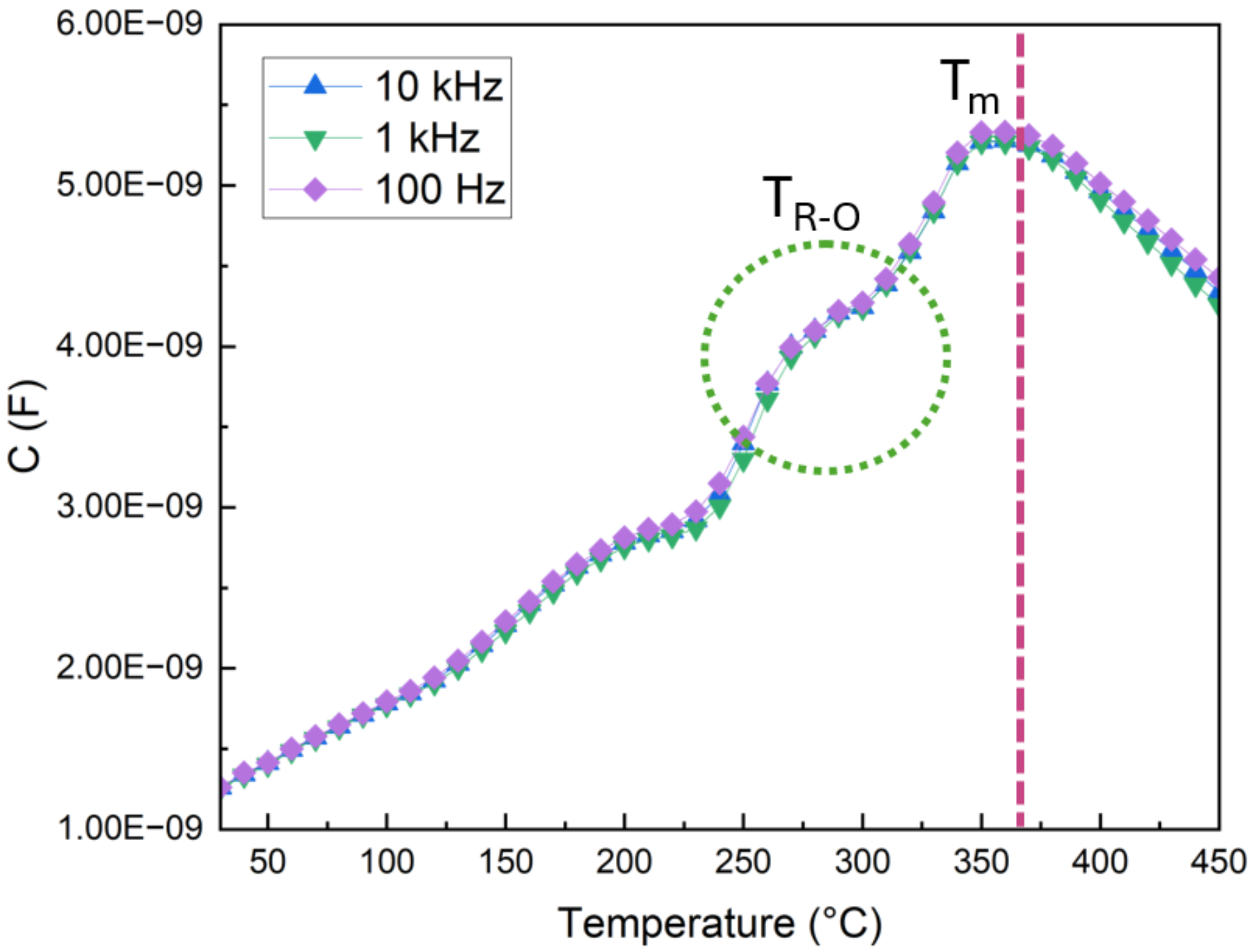

**Figure 10** – Temperature dependence of the capacitance extracted from impedance fitting using the RCPE model. $T_m$ is observed near 370 °C. The hump between 250 °C and 300 °C (circled in green) indicates the *R*3*c*+*Pnma* to pure *Pnma* transition ($T_{R\text{-}O}$).

### 2.3.3 Confirming the temperatures of transition

Three main structural transitions have therefore been identified: the transition from *R*3*c* + *P4bm* to *R*3*c* + *Pnma* at $T_d$, the transition from *R*3*c* + *Pnma* to *Pnma* at $T_{R\text{-}O}$, and from *Pnma* to a pseudo-cubic structure at $T_m$. Raman spectroscopy was employed to confirm these structural transitions and their critical temperatures. **Figure 11** shows the Raman spectra over the entire temperature range, with the modes divided into four regions. The region below 200 cm$^{-1}$ is attributed to the A-site cation-oxygen bonds. The second zone, between 200 cm$^{-1}$ and 425 cm$^{-1}$, corresponds to the Ti-O bonds, and the third zone, between 425 cm$^{-1}$ and 680 cm$^{-1}$, is related to $TiO_6$ octahedral rotations. The zone above 680 cm$^{-1}$ represents a superposition of acoustic and optical A1 and E modes. [50,51]

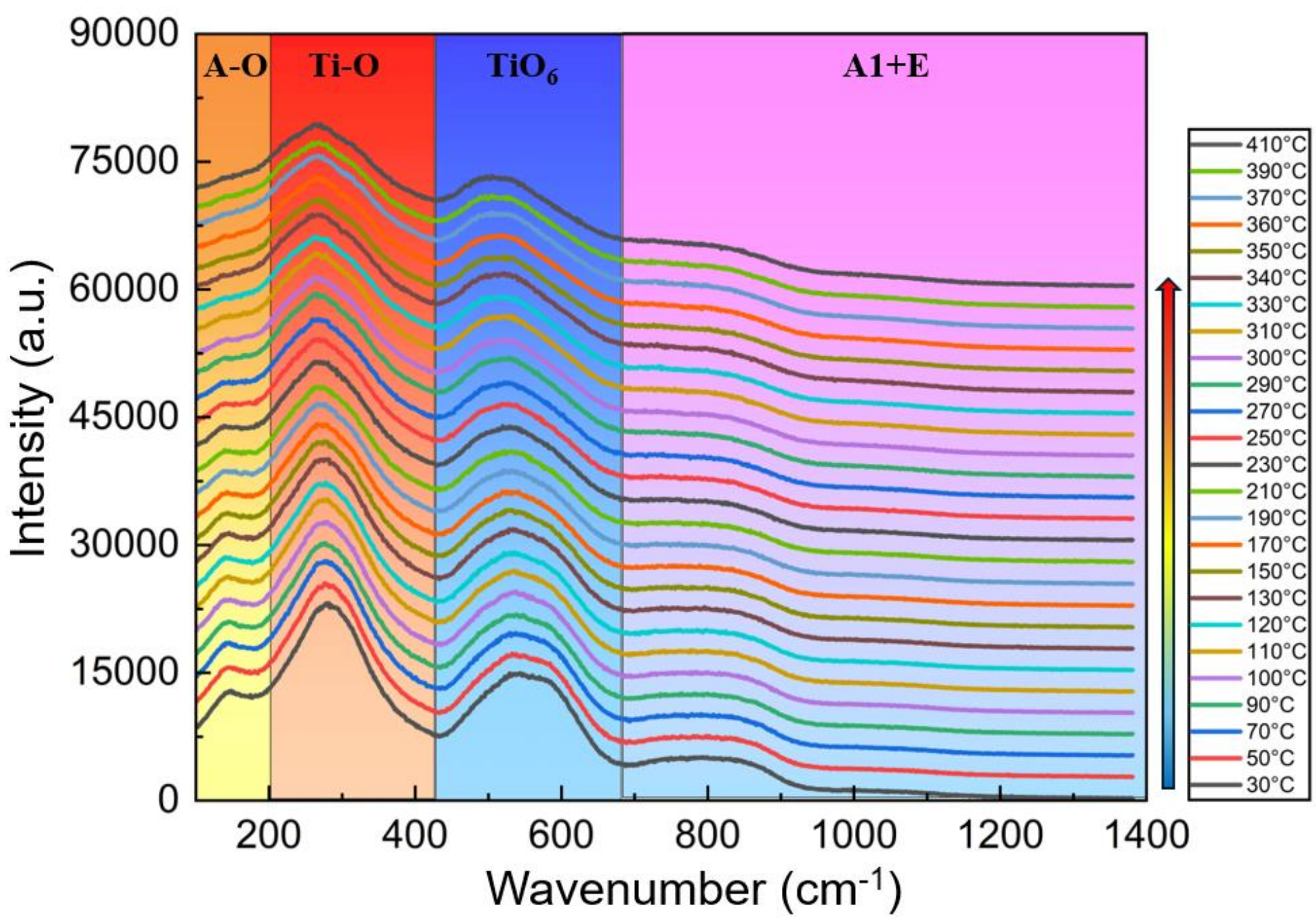

**Figure 11** – Temperature-dependent Raman spectra of BNT, divided into four regions: A–O vibrations (<200 cm$^{-1}$), Ti–O vibrations (200–425 cm$^{-1}$), $TiO_6$ octahedral rotations (425–680 cm$^{-1}$), and A1+E acoustic and optical modes above 680 cm$^{-1}$.

In addition to the seven modes usually needed for BNT, two additional modes were necessary to fit the Raman spectra (see **Figure S13**): the first corresponds to the $\nu_9$ mode at wavenumber above 1000 cm$^{-1}$, which is often omitted because the measurements stop before this range. The second is the $\nu_2$ mode, introduced to split the A-O bond vibration into two contributions to account for A-site disorder [52–54].

Several anomalies in the temperature evolution of the Raman-mode intensities are reported in **Figure S14**. The intensities were analyzed rather than the positions or widths because of the strong broadening of the modes, caused by structural disorder, high temperatures, and the coexistence of phases. These factors make it particularly difficult to reliably fit the peak positions. The first intensity anomaly occurs at 120 °C, corresponding to the point at which the bimodal distribution of coercive fields associated with the *P*4*bm* phase (**Figure S14.b**) becomes strongly deconvoluted. A second anomaly appears at 150 °C, which corresponds to $T_d$, *i.e.,* to the transition from *R*3*c* + *P*4*bm* to *R*3*c* + *Pnma*. A further discontinuity is observed near 290 °C, which is consistent with $T_{R\text{-}O}$, where the structure becomes purely orthorhombic (*Pnma*). Additional changes occur around 350 °C, corresponding to $T_m$, and a smaller anomaly appears near 200 °C that is not explained by the other measurements but might reflect the completion of the transition of peaks I2 and I3 toward the antiferroelectric state (**Figure 9.d**). Overall, the Raman analysis supports and consolidates the transitions identified using the other characterization techniques.

## 3. Conclusion

The combined study of the temperature-dependent symmetry with macroscopic dielectric response enabled the identification of several key features of BNT: at room temperature, the structure exhibits a coexistence of the rhombohedral *R*3*c* and tetragonal *P*4*bm* phases, as evidenced by XRD and electron diffraction, in agreement with the model proposed by Dorcet and Trolliard [12] that we completed with the addition of microstrain contribution in our Rietveld refinement. Analysis of the current-loop fitting reveals a splitting of the main current peak, indicating the presence of two ferroelectric contributions, one of which evolves into a bimodal coercive-field distribution at 80 °C. The depolarization temperature $T_d$ is 150 °C and corresponds to a structural transition from *R*3*c* + *P*4*bm* (where both phases are ferroelectric) to *R*3*c* + *Pnma* (where the latter is antiferroelectric). Between 250 °C and 300 °C, a further phase transition is detected by impedance spectroscopy and Raman measurements, corresponding to the disappearance of the *R*3*c* phase, in agreement with XRD results. Finally, the transition from the *Pnma* phase to a pseudo-cubic paraelectric phase at $T_m$ is observed around 350–370 °C using both impedance and Raman spectroscopy.

**Figure 12** summarizes the complete sequence of transitions with the different experimental techniques used to identify them. This work clearly demonstrates the need to combine structural and electrical analyses to understand the temperature-dependent behavior of complex dielectric ceramics.

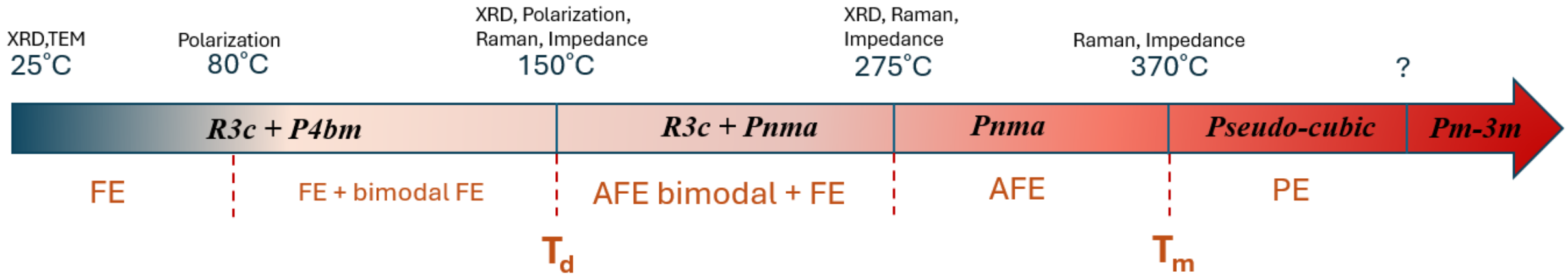

**Figure 12** – Schematic summary of the transition sequence in BNT identified in this study, showing the transition temperatures and the experimental techniques used to detect them. Crystallographic phases are indicated in black, while electrical orders are shown in red (FE: ferroelectric, AFE: antiferroelectric, PE: paraelectric), evidencing the necessity of combining structural and property characterization methods.

## 4. Experimental Section

The $Bi_{0.5}Na_{0.5}TiO_3$ (BNT) samples were synthesized using a gel-assisted citrate process. [55] The details of the synthesis process were presented in a previous work, [30] with the only difference being that the gel was calcined at 500°C for 6 h instead of 9 h. The same multi-step sintering process was used, as displayed in the previous work for the sintering of the pellets, since it provides the best properties for pure BNT. [30]

The phase purity was checked using a D8 Advance diffractometer (Bruker Inc., Germany) with a Cu Kα radiation. Patterns were recorded with a step of 0.02° (2θ) and a time of 2 seconds per step. The temperature-dependent X-ray diffraction (XRD) study was performed between 25 °C and 600 °C using a HTK1200N furnace (Anton Paar, Austria). XRD refinements were performed using Fullprof software [56] and the Rietveld refinement method. [57] Microstructure analyses and element dispersive X-ray (EDX) mapping were done with a scanning electron microscope JSM-7900F (JEOL, Japan) using an acceleration voltage of 10 keV. Prior to this, a thin layer of silver was sputtered onto the insulating sample to prevent electron charging at its surface. The average grain size was measured using imageJ software. [58] A room temperature transmission electron microscopy TEM study was performed with a JEM-2100F (JEOL, Japan) and X-Ray Photo-electron Spectroscopy (XPS) measurements were carried out with an Escalab Xi+ (Thermo Fisher Scientific, U.S.A.). Temperature dependent Raman measurements were performed from room temperature to 410°C with a Linkam stage coupled to a Renishaw InVia spectrometer using an excitation wavelength of 514.5 nm.

To measure the ferroelectric and dielectric properties, the pellets were polished to a thickness of 0.5 mm. They were then placed in a furnace at 400 °C for 5 hours to relax the stress caused by polishing. Afterward, the pellets were coated with a 7.5 mm diameter gold electrode using a SC7620 sputtering system (Quorum, U.K.). Dielectric properties were measured with a MFIA impedance analyzer (Zurich Instruments, Switzerland) in a PZMPS-CHH sample holder (Nextron Micro Probe System, Korea) from 30 °C to 450 °C in 2 °C increments. Ferroelectric hysteresis cycles were measured with a TFAnalyzer 2000E (AixACCT, Germany). A sinusoidal waveform with a 10 Hz frequency was used. Measurements on the samples have been performed with an electric field value of 60 $kV.cm^{-1}$ in silicone oil from 30 °C to 200 °C.

Data obtained from Raman spectroscopy, impedance spectroscopy, and current loops from ferroelectric properties measurements were fitted using Python scripts thanks to the lmfit package. [59]

**Acknowledgements**
The authors acknowledge the financial support of the Région Centre-Val de Loire (grant number 2021-00144736 APR IR Mathiole). This work has also been supported by the CERTeM 2020 Program, with the financial support of the Regional Council Centre-Val de Loire (CERTEM 2020 201400096812) and co-funded by the European Union through the European Regional Development Fund (ERDF 2017-EX002247)

**Data Availability Statement:**

Data will be made available upon reasonable request.

# Supporting Information

**Consistent transition model for $Bi_{0.5}Na_{0.5}TiO_3$ from temperature-dependent structural and electrical properties**

*Thomas Fourgassie [1,2], Omar Ibder [2], Cosme Milesi-Brault [2], Anna Katharina Ott [1], Eric Bourhis [3], Pascal Andreazza [3], Pierre-Eymeric Janolin [2] and Cécile Autret-Lambert * [1,2]*

Affiliations:

[1] Laboratoire GREMAN UMR7347, University of Tours, Tours, France

[2] Laboratoire SPMS UMR8580, University Paris-Saclay, Gif-sur-Yvette, France

[3] ICMN UMR7347, University of Orléans, Orléans, France

*Corresponding author e-mail: autret@univ-tours.fr*

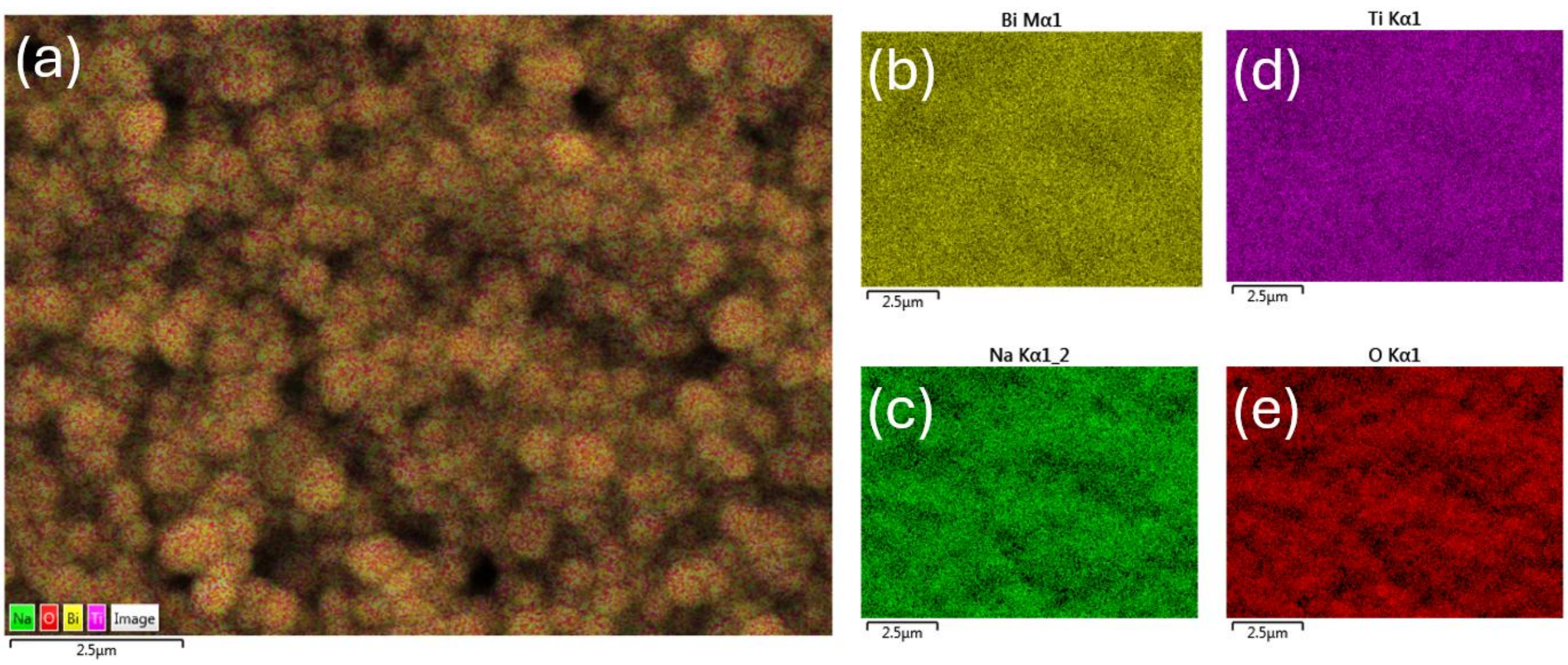

**Figure S1 –** EDX elemental mapping of BNT showing homogeneous chemical distribution: (a) combined map, (b) Bi, (c) Na, (d) Ti and (e) O

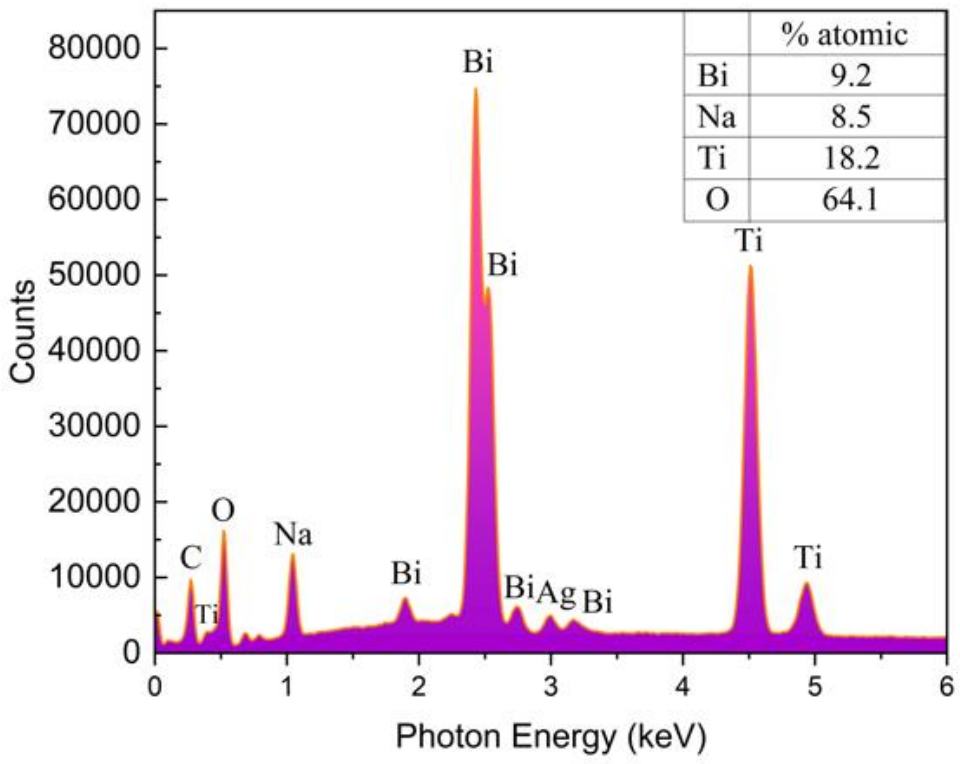

| | % atomic |
|---|---|
| Bi | 9.2 |
| Na | 8.5 |
| Ti | 18.2 |
| O | 64.1 |

**Figure S2 –** Photon counts at different energies during the EDX analysis with the atomic percentage of each element that shows a chemical composition close to the targeted composition for BNT. The peaks attributed to silver are due to the thin silver deposition made on the sample to avoid electron charging on the surface of the sample.

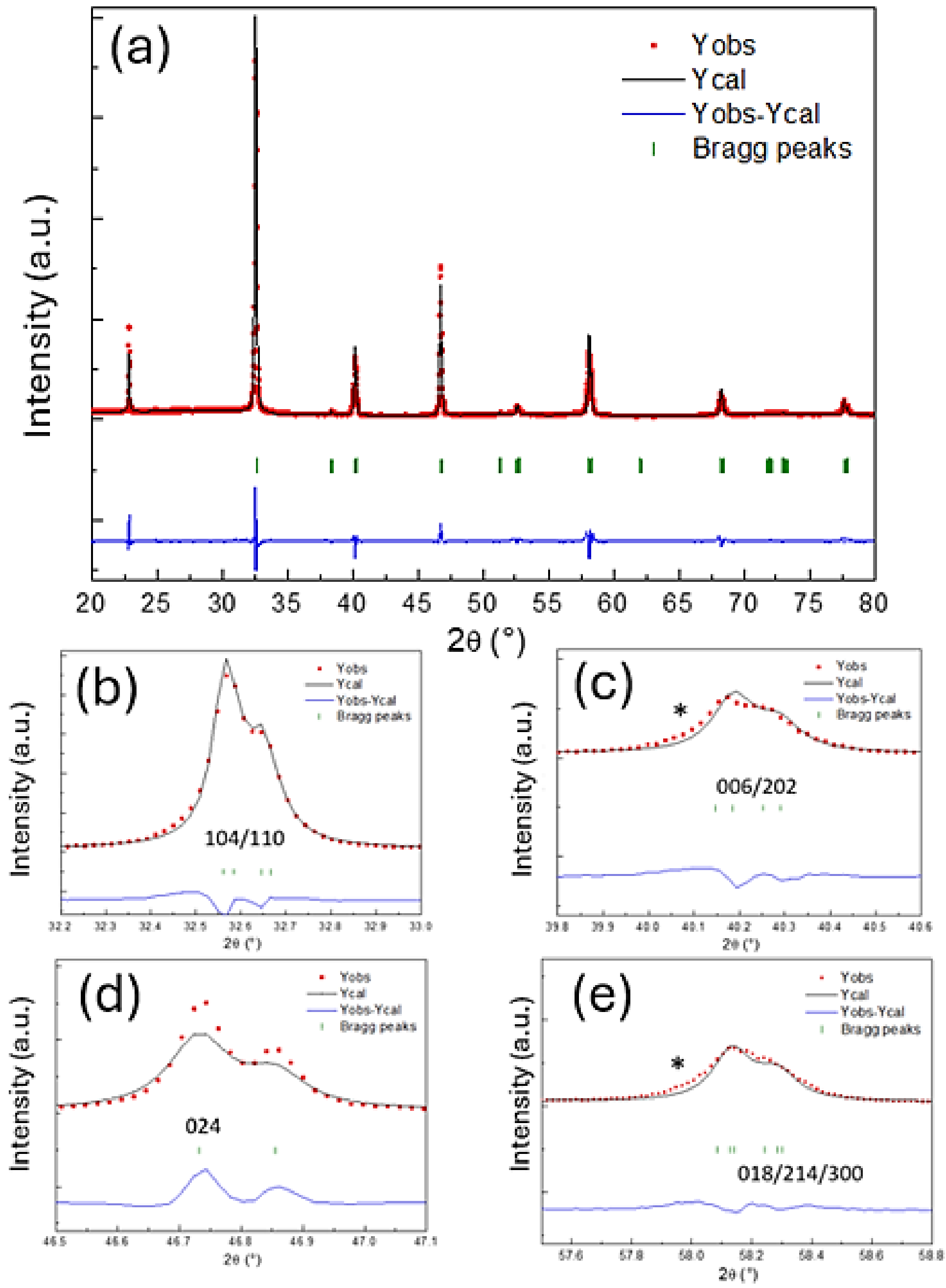

**Figure S3 –** (a) Rietveld refinement of the room temperature XRD pattern using a single *R*3*c* phase. Enlarged views of important peaks : (b) $(104)_H$ and $(110)_H$, (c) $(006)_H$ and $(202)_H$, (d) $(024)_H$ and (e) the $(018)_H$, $(214)_H$ and $(300)_H$. Star symbols peak shoulders not accounted for by the pure *R*3*c* phase.

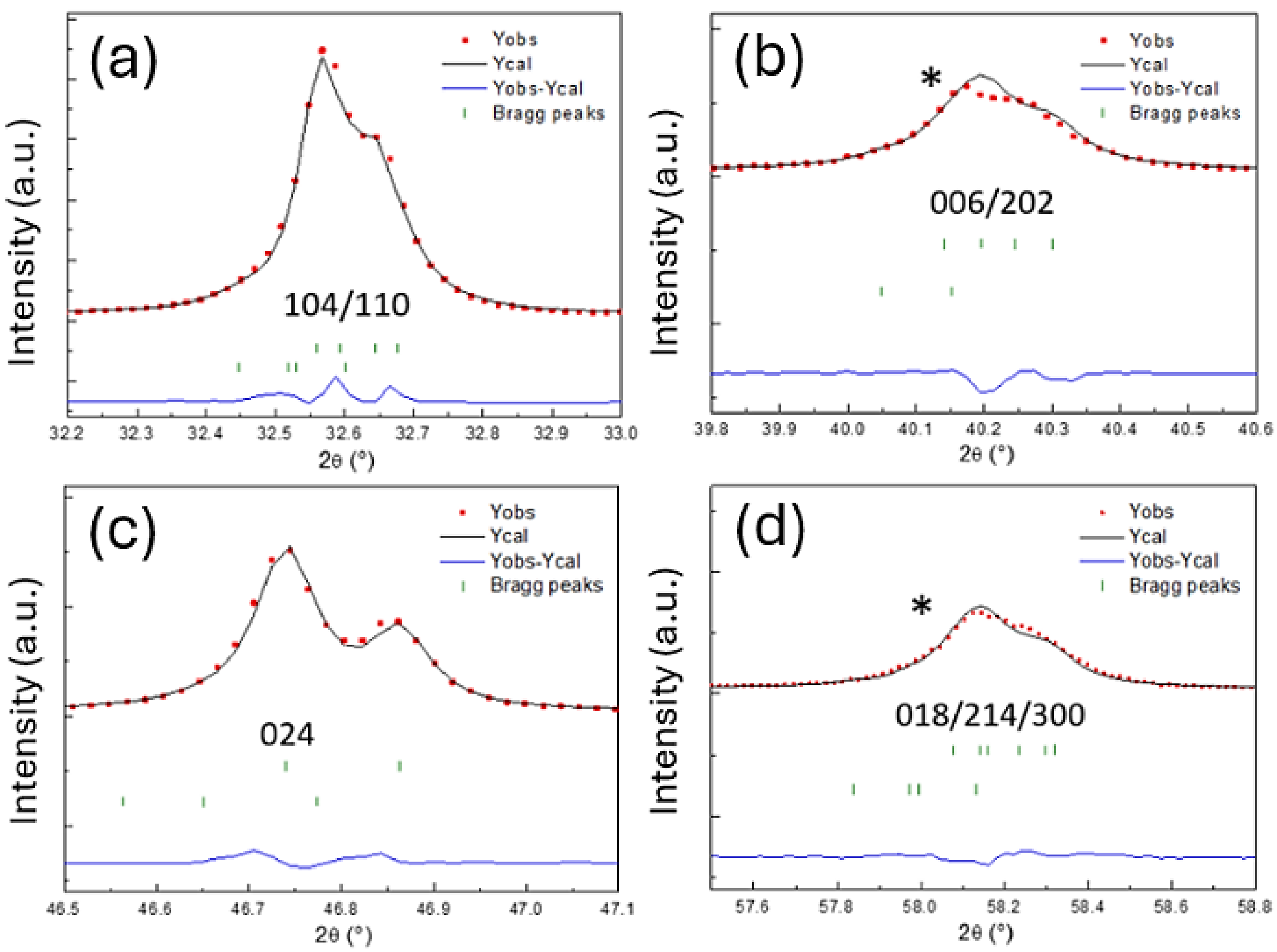

**Figure S4** – Enlarged views of the room temperature XRD scan refinement using a mixed *R*3*c* and *P*4*bm* phases : (a) $(104)_H$ and $(110)_H$, (b) $(006)_H$ and $(202)_H$, (c) $(024)_H$ and (d) $(018)_H$, $(214)_H$ and $(300)_H$. Star symbols highlight peak shoulders now well fitted with the *P*4*bm* phase.

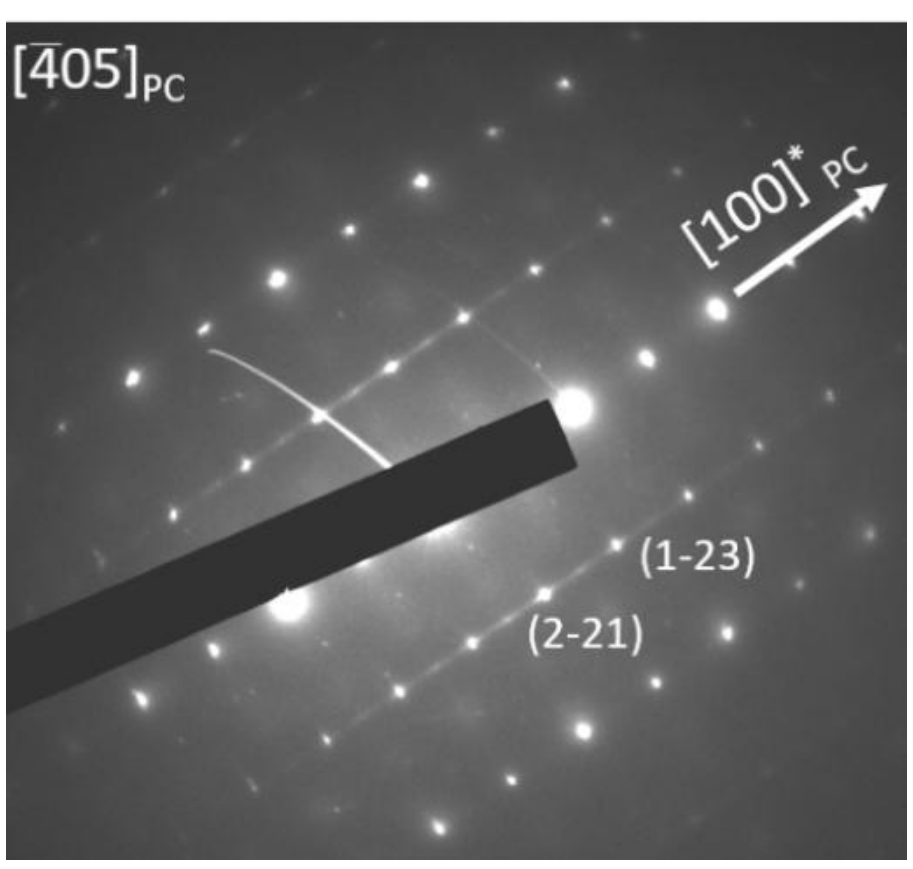

**Figure S5** – Electron diffraction pattern along the $[\bar{4}05]_{PC}$ zone axis showing diffuse streaks parallel to $[100]_{PC}$ attesting for a high density of defect along this direction.

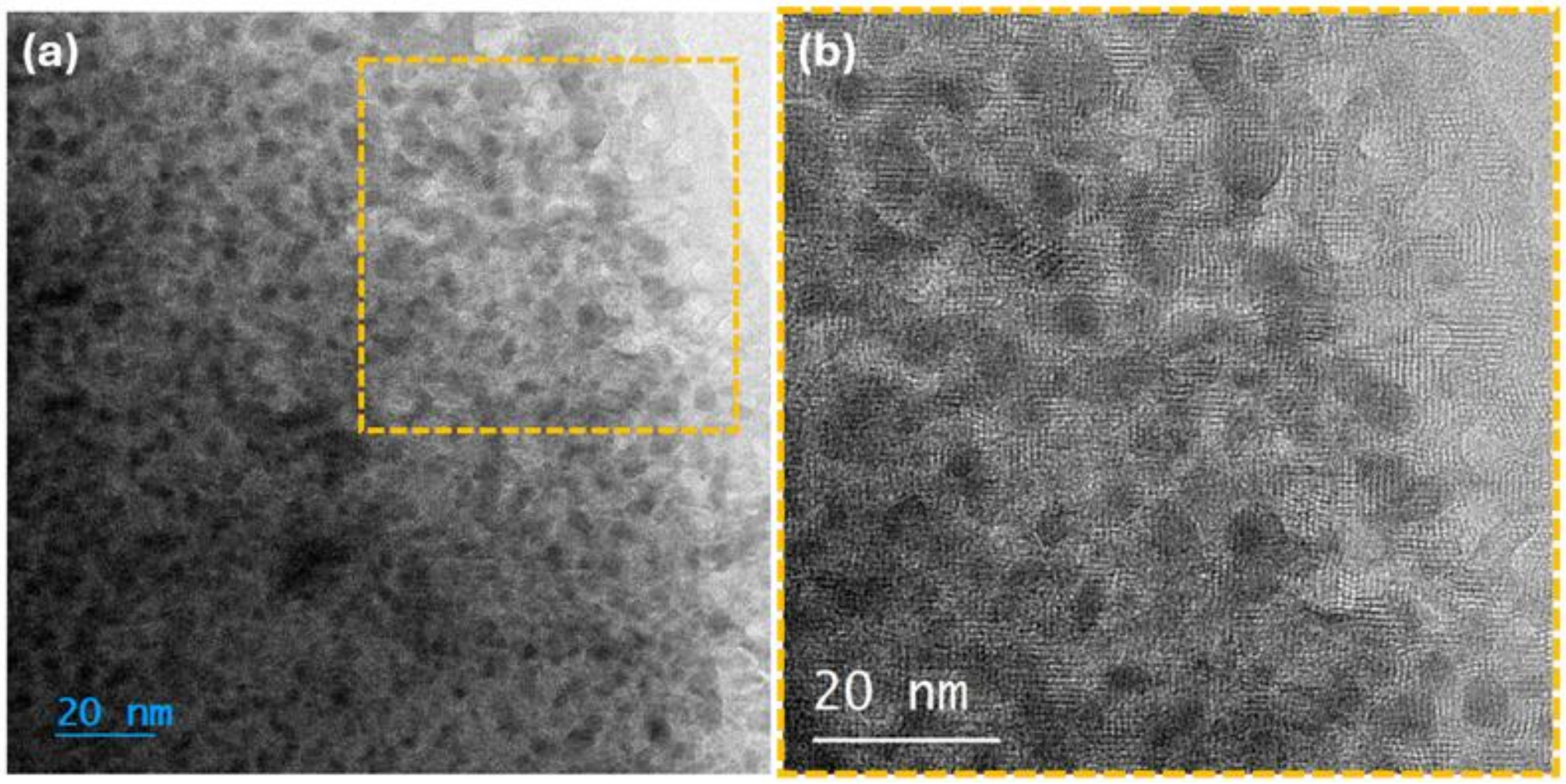

**Figure S6 –** (a) HRTEM image after prolonged electron beam exposure and (b) enlarged view (Orange Square) showing the clear contrast between nanodomains.

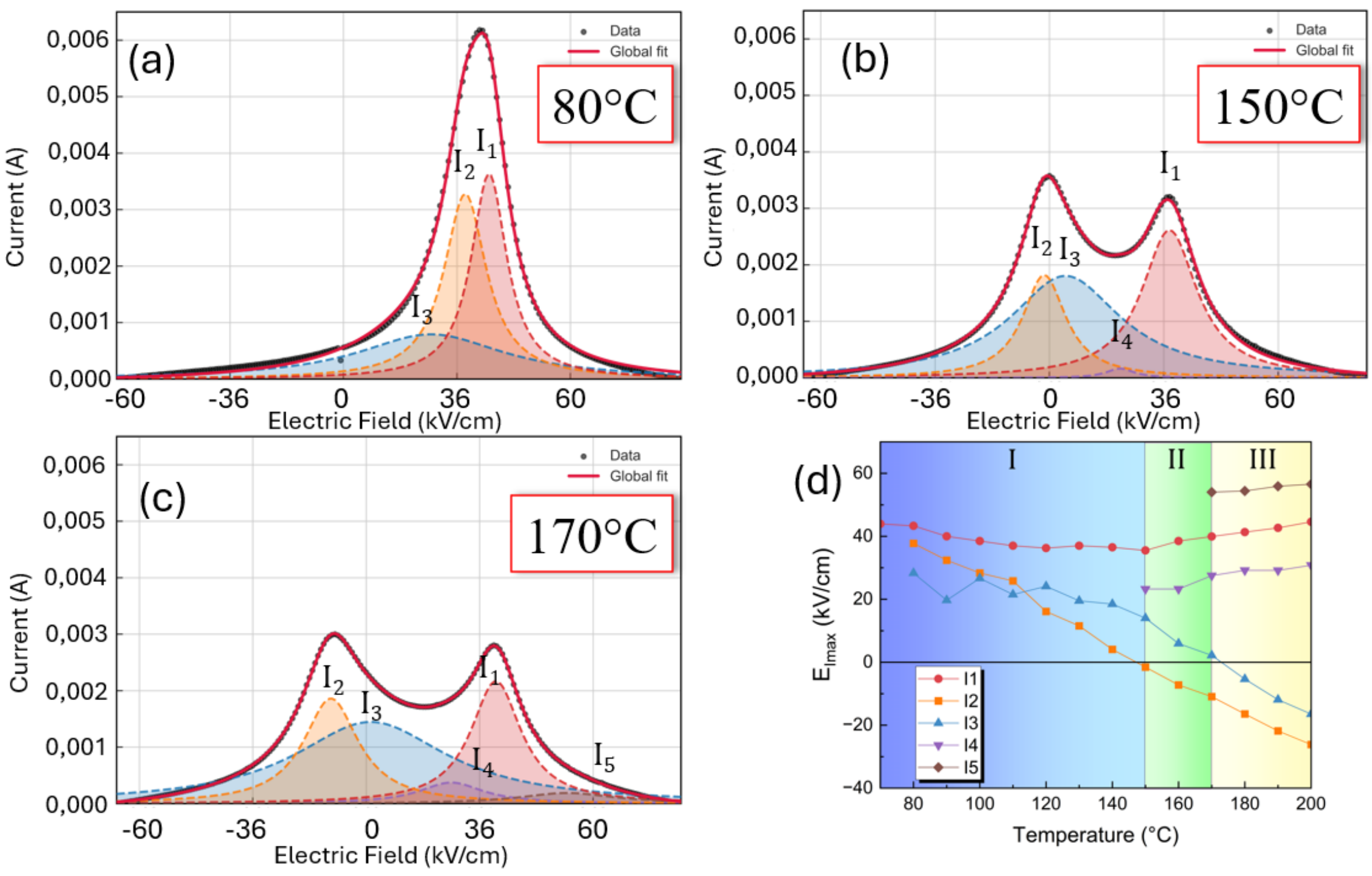

**Figure S7 -** Fitting of current–electric field curves using Lorentzian functions: (a) three peaks at 80 °C, (b) four peaks at 150 °C, (c) five peaks at 170 °C, (d) and temperature evolution of peak positions highlighting three distinct regimes.

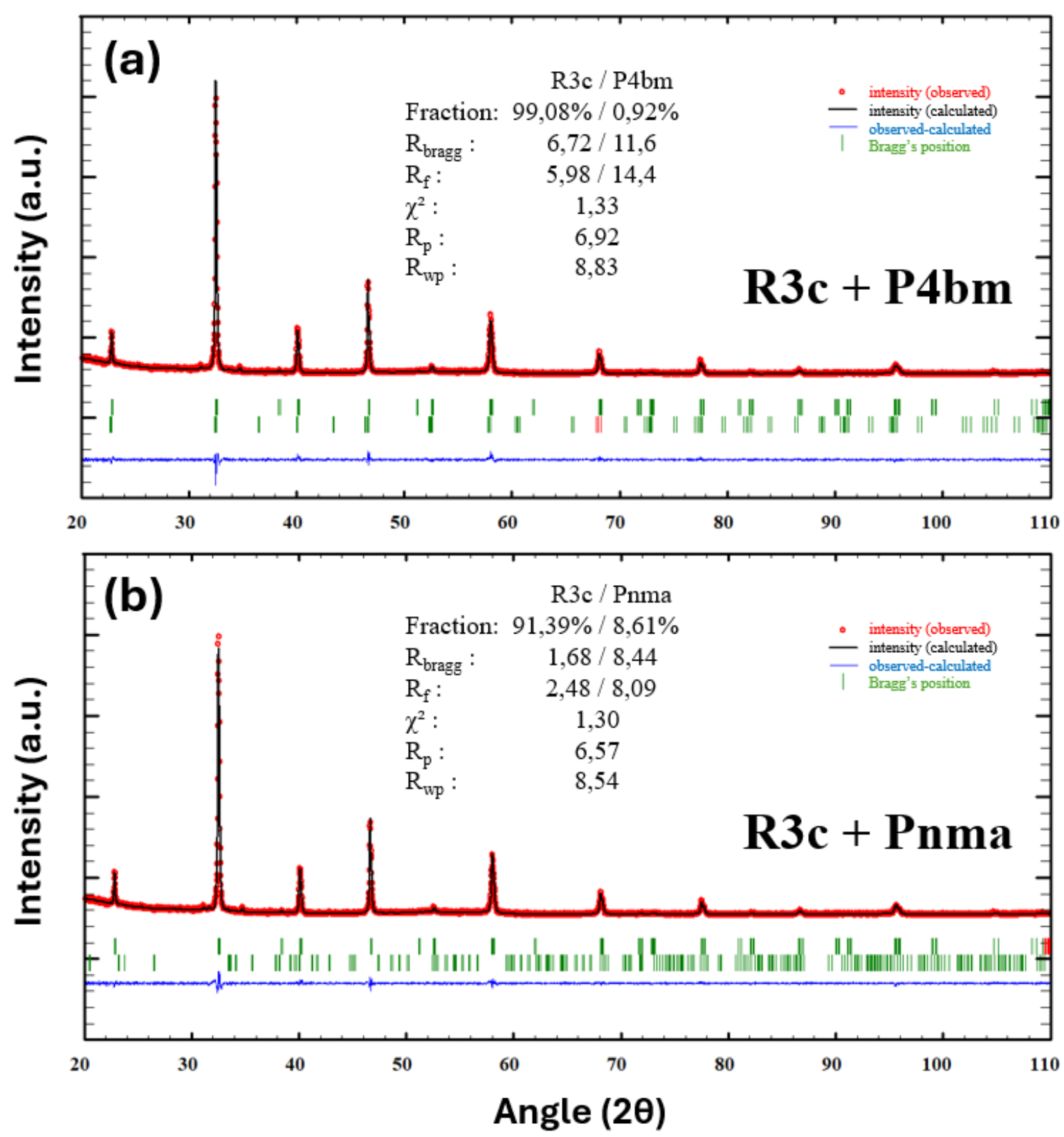

**Figure S8 –** Rietveld refinement of XRD data at 175 °C using (a) *R*3*c* and *P*4*bm*, and (b) *R*3*c* and *Pnma* structural models. Both fits are satisfactory, with improved reliability factors for the R3c + Pnma model.

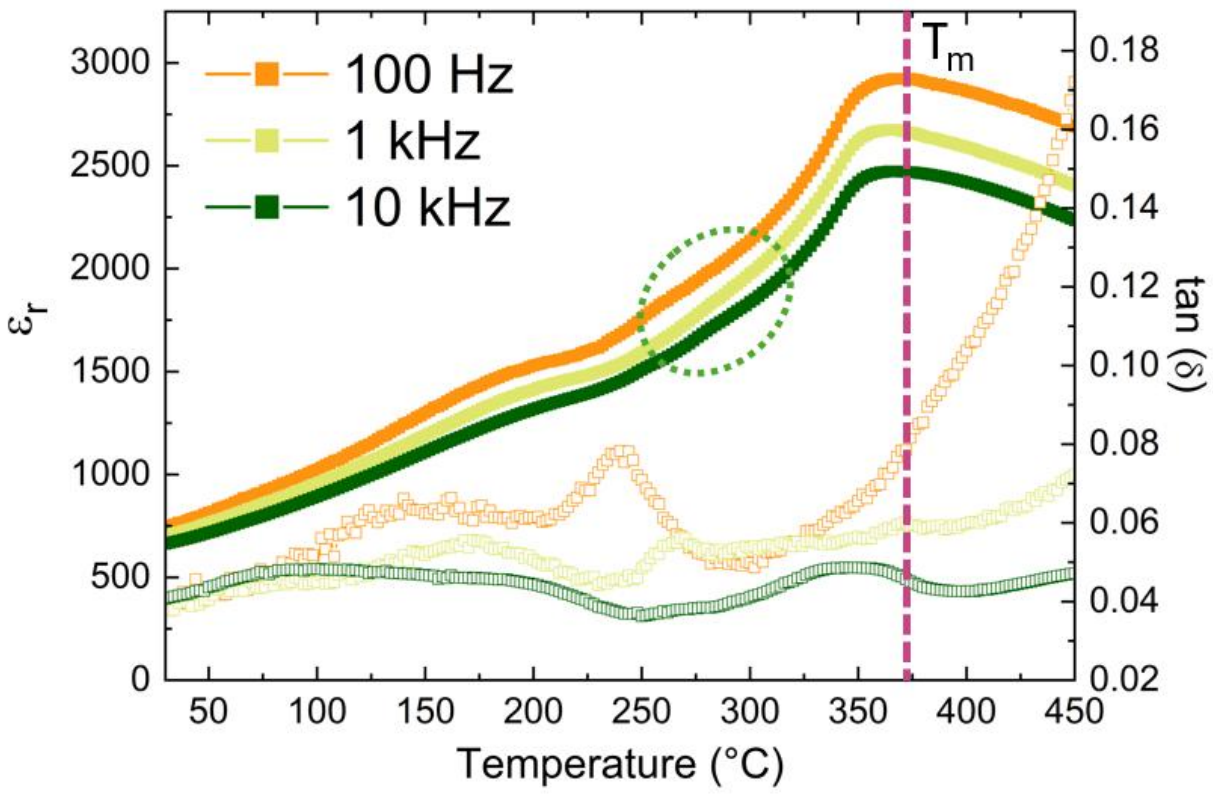

**Figure S9 –** Relative permittivity (filled symbols) and dielectric losses (open symbols) of BNT as a function of temperature and frequency. $T_m$ is observed at around 370 °C and the green circle indicates an anomalous dielectric response.

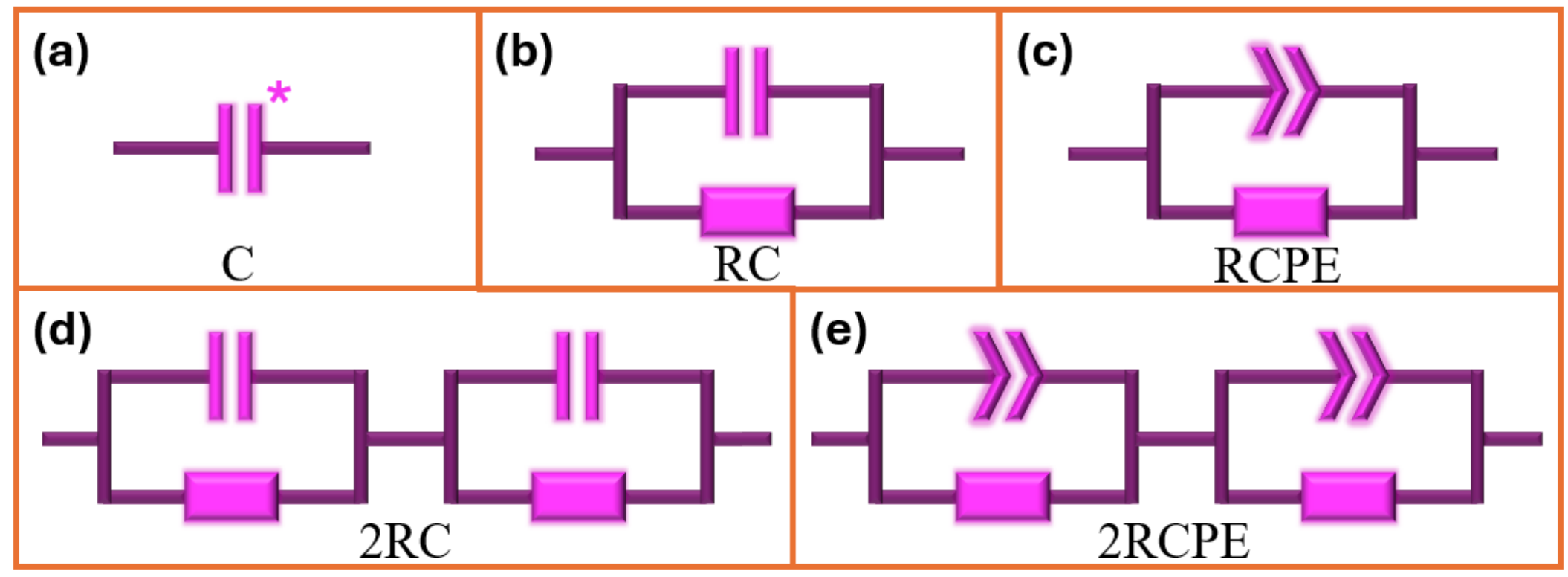

**Figure S10** – Schematic representation of the equivalent circuits used for impedance fitting: (a) C, (b) RC, (c) RCPE (ZARC), (d) 2RC, and (e) 2RCPE models.

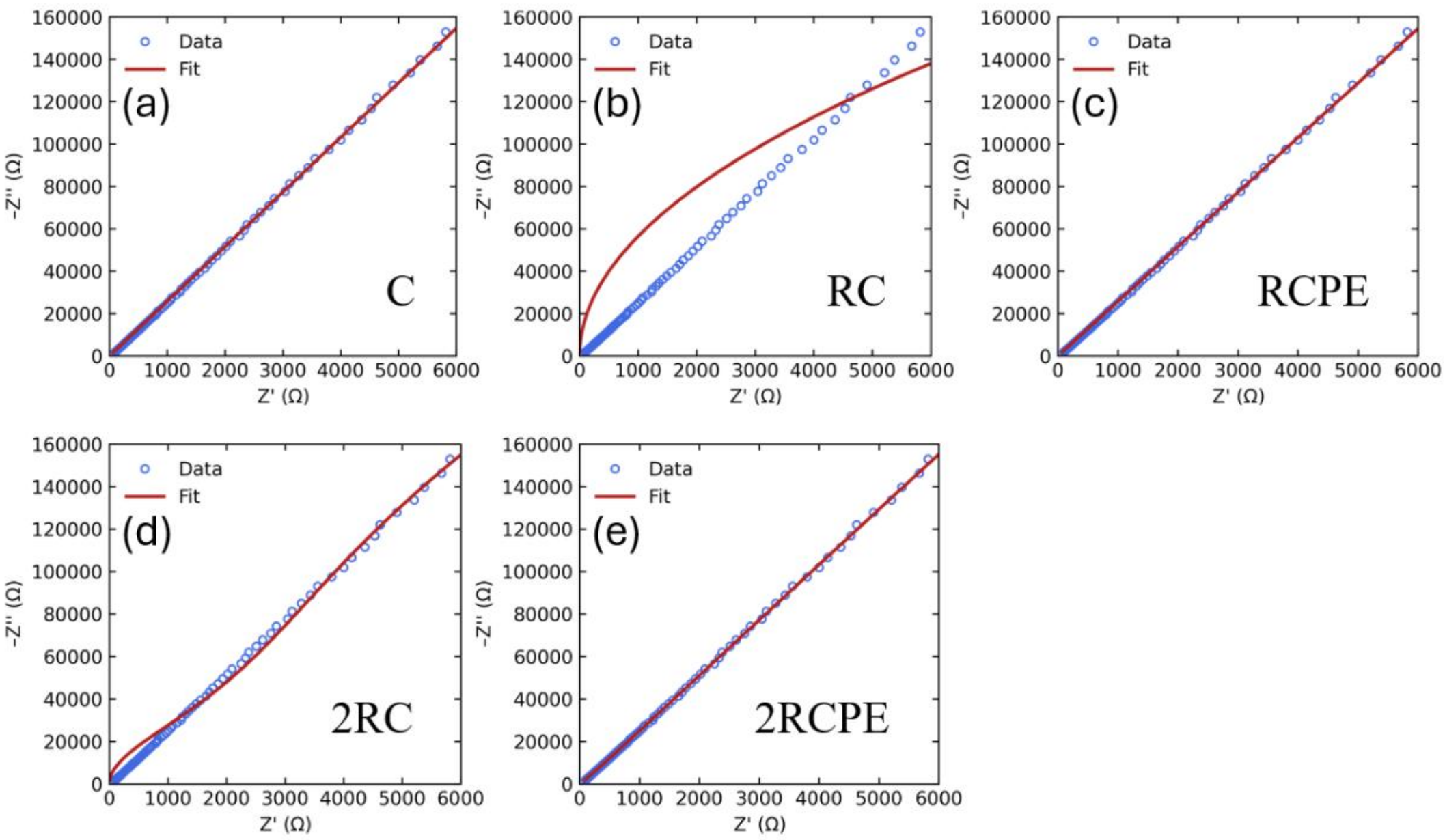

**Figure S11** – Nyquist plots comparing experimental impedance data (open symbols) and fits (red lines) for the (a) C, (b) RC, (c) RCPE, (d) 2RC, and (e) 2RCPE models.

At this stage, three models (C, RCPE, and 2RCPE) provide a satisfactory fit of the impedance data. To discriminate between them, the fitting residuals were examined (**Figure S12**). These residuals clearly show that the C model can be discarded, while both the RCPE and 2RCPE models offer good agreement with the data, with a slightly better fit coming from the 2RCPE model.

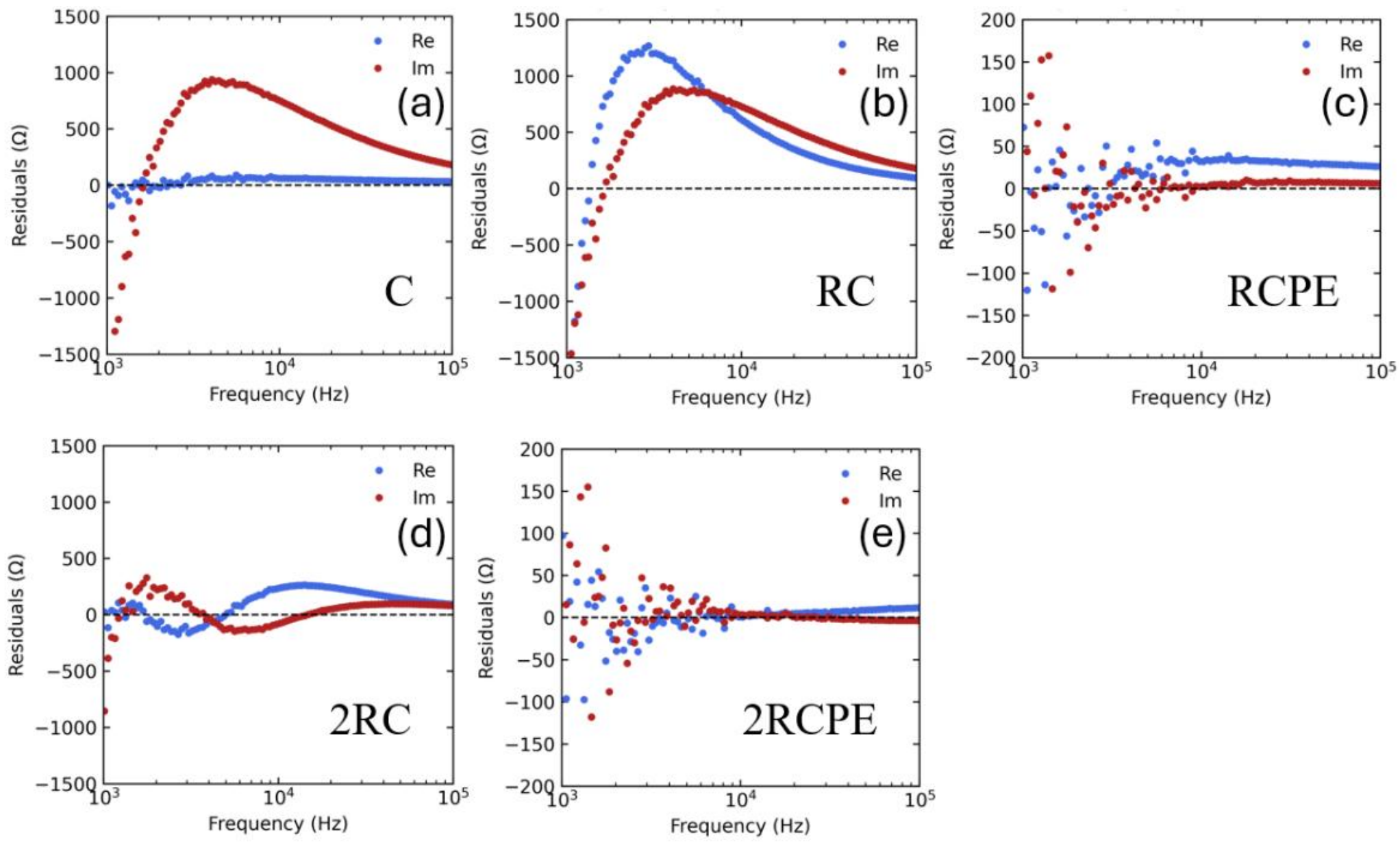

**Figure S12 –** Residuals of the real (blue) and imaginary (red) parts of the impedance for the different equivalent-circuit models, showing improved agreement for the RCPE and 2RCPE models.

Finally, to determine the most suitable model, the Akaike Information Criterion (AIC) and the Bayesian Information Criterion (BIC), defined by **equations S1 and S2**, were examined. In these expressions, n is the dataset size, k is the number of adjustable parameters, and L is the maximum of the likelihood function. Since the 2RCPE model has more adjustable parameters than the RCPE model, these criteria allow evaluation of the relationship between model complexities and fitting accuracy. The values obtained for both criteria are reported in **Table S1** in the supporting information.

$$AIC = 2k - 2ln(L) \qquad (S1)$$

$$BIC = k\, ln(n) - 2ln(L) \qquad (S2)$$

When examining these information criteria (**Table S1**), the absolute values have no intrinsic meaning, as they are intended only for model comparison. Therefore, only the differences between values are relevant. The model with the lowest value for a given criterion corresponds to the smallest information loss and is considered the most suitable according to that criterion. In our case, the RCPE model appears to be the most appropriate, providing the best compromise between fitting accuracy and model complexity according to both criteria.

| Equivalent Circuit | AIC | BIC |
|---|---|---|
| C | 6504 | 6512 |
| RC | 6757 | 6765 |
| RCPE | 5294 | 5306 |
| 2RC | 6525 | 6541 |
| 2RCPE | 6200 | 6224 |

**Table S1 –** Akaike (AIC) and Bayesian (BIC) information criteria obtained for each equivalent-circuit model. Minimum values (highlighted) indicate the RCPE model as the most appropriate.

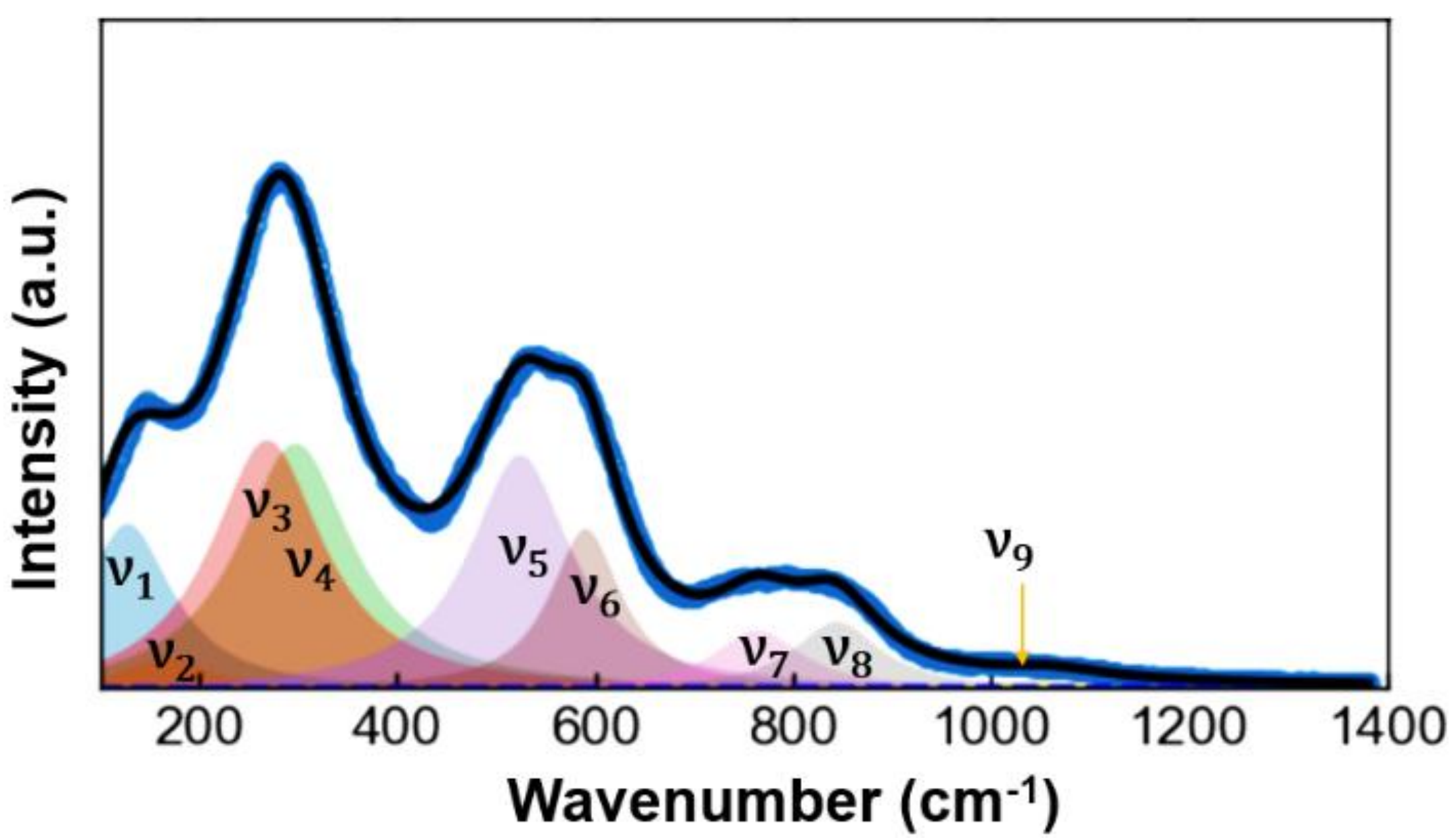

**Figure S13 –** Example of Raman spectrum fitting at 30 °C showing the 9 vibrational modes used in the analysis.

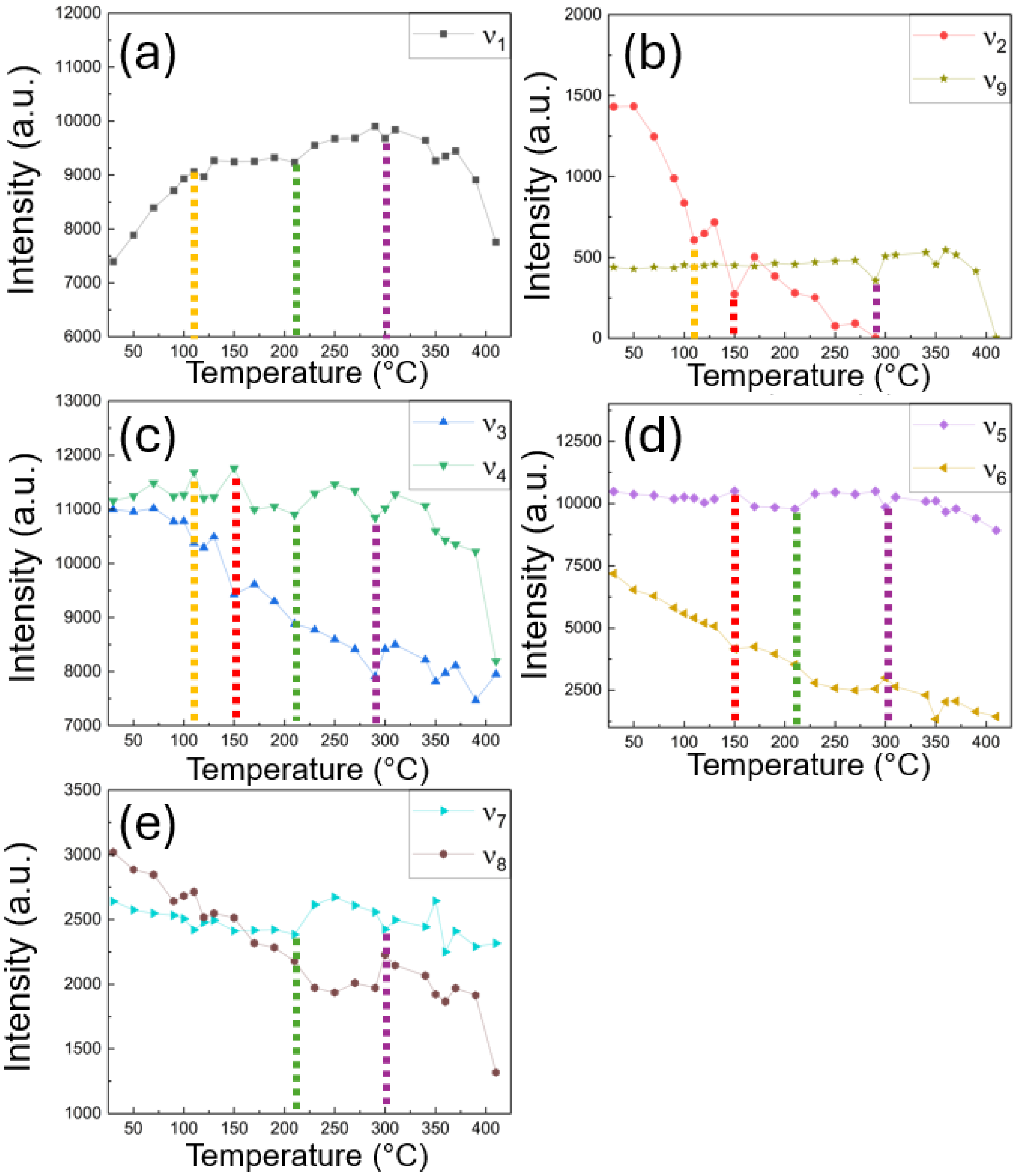

**Figure S14** – Temperature evolution of Raman-mode intensities, showing anomalies near 120 °C (yellow line), 150 °C (red line), 200 °C (green line) and 300 °C (purple line), as well as increased instability above 350 °C, consistent with transitions identified by other techniques.